\documentclass[10pt,journal,compsoc]{IEEEtran}
\usepackage[cmex10]{amsmath}
\usepackage{amssymb}
\usepackage{graphicx} 
\usepackage{url}
\usepackage{algorithm}
\usepackage{algorithmic}
\usepackage{color}
\usepackage{changepage}
\newtheorem{algorisme}{Algorithm}{\bfseries}{\itshape}

\newtheorem{proposition}{Proposition}{\bfseries}{\itshape}
{\bfseries}{\itshape}
\begin{document}

\title{General Confidentiality and Utility 
Metrics for Privacy-Preserving 
Data Publishing Based on the Permutation Model}

\author{Josep Domingo-Ferrer,~\IEEEmembership{Fellow, IEEE,} 
Krishnamurty Muralidhar and Maria Bras-Amor\'os%
\thanks{J. Domingo-Ferrer and Maria Bras-Amor\'os are 
with the UNESCO Chair in Data Privacy, 
CYBERCAT-Center for Cybersecurity Research of Catalonia,
Department of Computer Engineering and Mathematics,
Universitat Rovira i Virgili,
Av. Pa\"{\i}sos Catalans 26, E-43007 Tarragona, Catalonia,
e-mail \{josep.domingo,maria.bras\}@urv.cat.
\newline K. Muralidhar is with the Price College of Business,
University of Oklahoma,
307 West Brooks, Adams Hall Room 10, Norman OK 73019-4007, USA,
e-mail krishm@ou.edu.}}%

%\markboth{IEEE TRANSACTIONS ON DEPENDABLE AND SECURE COMPUTING,~Vol.~?, No.~?, Month~YYYY}%
%{J. Domingo-Ferrer \MakeLowercase{\textit{et al.}}: General Confidentiality 
%and Utility Metrics}

\IEEEtitleabstractindextext{%
\begin{abstract}
%JOSEP11. Rewritten: anonymization aka SDC.
Anonymization for privacy-preserving data publishing, also known
as statistical disclosure control (SDC), can be viewed under the lens
of the permutation model. According to this model, any SDC method
for individual data records is functionally equivalent to 
a permutation step plus a noise addition step, where the 
noise added is marginal, in the sense that it does not 
alter ranks.
Here, we propose metrics to quantify the data confidentiality and
 utility 
achieved by SDC methods based on the permutation 
model. 
%JOSEP3. Modified to distinguish privacy and confidentiality.
%JOSEP11. Changed due to the comments of one reviewer to the second
%revision.
We distinguish two privacy notions: 
in our work, anonymity refers to subjects and hence mainly to protection
against record re-identification, 
whereas confidentiality refers to the 
protection afforded to attribute values against attribute disclosure.
Thus, our confidentiality metrics are useful even if 
using a privacy model ensuring an anonymity level {\em ex ante}.
The utility metric is a general-purpose metric that 
can be conveniently traded off against 
the confidentiality metrics, because all of 
them are bounded between 0 and 1. As an application,
we compare the utility-confidentiality trade-offs achieved
by several anonymization approaches, including 
privacy models ($k$-anonymity and $\epsilon$-differential privacy)
as well as 
 SDC methods (additive noise,
multiplicative noise and synthetic data) used without 
privacy models.
\end{abstract}

}

\maketitle

\IEEEpeerreviewmaketitle

\section{Introduction}

%JOSEP3. Reviewer 2. I have extended the intro.
\IEEEPARstart{S}{ince} the turn of the century, we are fully immersed 
in the information society. Most human activities leave digital traces 
that someone collects and stores. Social media, the internet of things,
bank transactions, purchases at stores are just a few ways 
of gathering data on people.  
Such a massive data collection has many advantages:
increased business opportunities, better and more rigorous
research and, in general, rosier prospects of 
 improving the well-being of the human race. 

Yet, accumulating, sharing and publishing 
personally-identifiable information (PII) 
has also a disquieting side,
as it invades the privacy of the subjects to whom PII relate;
a famous example is the teenager pregnancy guess reported in~\cite{teenager}.
Data protection legislation, epitomized by the EU General Data Protection
Regulation~\cite{gdpr}, tries to protect citizens by restricting
the accumulation of PII. 
Anonymizing PII, {\em i.e.} turning them into data that are not 
personally identifiable but still retain substantial analytical
utility, is a way to enable data analysis, sharing and even publishing
without violating the data protection laws.

Anonymization for privacy-preserving data publishing is 
also known as statistical disclosure control 
(SDC, \cite{Hundepool12}).
The usual setting in          
anonymization is for a {\em data controller} (the entity that 
manages and releases the data, and often owns them) 
to hold the original data (with the original responses by the {\em subjects}) 
and modify them to reduce the disclosure risk.
Then the controller publishes the 
anonymized data or shares them with {\em users}, typically 
data analysts or researchers ---who expect the anonymized
data to be still useful.
It may occur that some of the users   
behave as {\em intruders} and try to perform disclosure attacks 
on the anonymized data. Disclosure can be of two types:
\begin{itemize}
\item {\em Re-identification disclosure}, whereby the intruder determines
 the subject to whom an anonymized data item corresponds;
\item {\em Attribute disclosure}, in which the 
anonymized data help the intruder
 to estimate 
the value of a confidential attribute for a certain subject.
\end{itemize}

Data at the individual level, such that each record corresponds
to one individual subject (person, enterprise, etc.), 
are known as microdata. From microdata, other formats can be 
derived, such as tables (the traditional output of national 
statistical institutes) and on-line queryable databases
(that answer statistical queries on an underlying microdata set).
Here, we will focus on microdata. 
%JOSEP3. End of extension.

%JOSEP3. Reviewer 2. Changed passive sentences into active.
The traditional approach to anonymization, still dominant among statistical
agencies, can be called ``utility-first'': 
the controller runs an SDC method~\cite{Hundepool12}
with a heuristic parameter choice and with suitable utility preservation
properties on the microdata set. After that, the controller 
measures the risk of disclosure, which she can do 
empirically by attempting record linkage 
between the original and the anonymized data sets~\cite{Torra03},
or analytically by using generic metrics ({\em e.g.}~\cite{Lambert93}) 
or metrics tailored to a specific SDC method 
({\em e.g.}~\cite{Elamir06} for 
 sampling-based SDC). If the controller
deems the remaining risk too high, she re-runs
the anonymization method with more confidentiality-stringent 
parameters and probably more utility sacrifice.

Whereas most utility-first SDC methods obtain each anonymized record
by masking a certain original record, synthetic data are an exception.
In this case, the 
anonymized data set consists of synthetic/simulated data that preserve
a set of utility characteristics of the original data set.
Since there is no direct mapping between original and synthetic
records, synthetic data are often regarded as the safest utility-first
approach. Unfortunately, this lack of mapping 
also makes it difficult to quantify 
the confidentiality actually achieved, because many confidentiality
metrics need to compare each anonymized record with its corresponding
original record.

An alternative anonymization approach can be termed ``privacy-first''
and is based on the notion of {\em privacy model}, which is a 
condition dependent on a parameter that guarantees an upper bound 
on the risk of reidentification disclosure and perhaps also on 
the risk of attribute disclosure by an intruder. Well-known
privacy models include $k$-anonymity~\cite{Samarati98} and 
its extensions, as well as $\epsilon$-differential privacy~\cite{Dwork06}.
%JOSEP3. Reviewer 2. Passive sentences into active.
The controller can enforce
a certain privacy model using one or several 
specific SDC methods whose
parameters are a function of the model parameters.
For example, $k$-anonymity can be attained with a combination
of generalization and suppression or with microaggregation~\cite{Domingo05};
$\epsilon$-differential privacy is normally attained via 
noise addition.
%JOSEP3. Reviewer 2. Passive into active.
There may be two issues with the privacy-first approach: on the one
side, if the controller chooses too stringent a parameter for the privacy 
model,
the utility of the anonymized data may be too low; on the other side,
if she chooses too relaxed a parameter, the protection 
given by the privacy model may be insufficient.

Thus, no matter whether the controller
follows the utility-first or the privacy-first
approaches, she needs to measure the 
 utility and the protection provided by a certain anonymization method.
However, SDC methods for microdata rely on a diversity of principles~\cite{Hundepool12},
and this makes it difficult to analytically compare
their utility and data protection properties~\cite{Duncan91}; 
this is why one usually resorts to empirical 
comparisons~\cite{Domingo-Ferrer01}.

\subsection*{Contribution and plan of this paper}

%JOSEP3. Reviewers 1 and 2. Distinction between confidentiality and utility.
%JOSEP11. Changed 
In this paper, we
present new {\em confidentiality} and {\em utility} 
metrics for anonymized data.
Let us briefly define these notions.
We use utility in the customary sense of preserving
the statistical properties of the original data. 
On the other hand, confidentiality is one of the two 
main privacy notions in statistical disclosure control,
the other being anonymity. 
Whereas anonymity refers to subjects and hence mainly to protection
 against record re-identification, confidentiality 
refers to the protection afforded to attribute values 
against attribute disclosure.

Specifically, we exploit the unified view of anonymization 
afforded by the permutation model~\cite{Domingomuralidhar16} to derive 
 bounded confidentiality metrics for microdata anonymization that are
 based on the 
relative amounts of permutation undergone by the different
attributes of a data set.
We then give a bounded utility metric that can be used to 
evaluate the trade-off between utility and confidentiality 
and also to compare this trade-off among different 
utility-first SDC methods as well as among different privacy
models.

In Section~\ref{background}, we give background on 
%JOSEP11.
anonymity vs confidentiality, on
the permutation
model and on canonical correlation, a primitive
that we will use to construct our confidentiality metrics. 
Section~\ref{discussion} makes the case for using 
permutation to assess confidentiality and utility.
%In Section~\ref{covariance} we make the case for evaluating
%permutation and hence privacy and utility based on covariance matrices.
In Section~\ref{privacy} we present
the confidentiality metrics and in Section~\ref{utility} we present
the utility metric.
Empirical work on the operation of the new metrics 
is described in Section~\ref{empirical}; 
we compare the utility-confidentiality trade-offs attained 
by several anonymization approaches, including   
privacy
models ($k$-anonymity and differential privacy)
and utility-first SDC methods (additive noise,
multiplicative noise and synthetic data).
Section~\ref{related} reviews related work.
Finally, conclusions and lines of future research 
are summarized in Section~\ref{conclusions}.

\section{Background}
\label{background}

%JOSEP3. Reviewers 1 and 2. Added section.
\subsection{Anonymity vs confidentiality}
\label{privconf}

%JOSEP11.
As mentioned above, we use anonymity to refer to protection
against record re-identification and confidentiality 
to protection against attribute disclosure. 
%JOSEP11.
In fact, a given anonymity level 
 can co-exist with different levels of confidentiality:
\begin{itemize}
\item In $k$-anonymity, if the original data set happens
to be already $k$-anonymous (this is quite unlikely, but possible),
then no SDC masking is needed for anonymity. Even though the anonymity level 
is $k$, confidentiality is zero, because the original attribute
values are not modified. On the other hand, if attributes 
need to be heavily masked to attain $k$-anonymity, confidentality 
is in general nonzero.
\item In $\epsilon$-differential privacy, if the original data set or query
have very low sensitivity (they depend very little on the 
absence or presence of any single subject in the data), 
very little noise needs to be added to satisfy the model.
Thus in this case, anonymity is inversely proportional to $\epsilon$, 
but confidentiality is
low. Conversely, if sensitivity is high, a lot of noise is added
and confidentiality is high even though 
$\epsilon$ is the same. 
\item No matter whether $k$-anonymity or $\epsilon$-differential privacy
are used, for the same anonymity level 
confidentiality can be expected to grow with the number 
of attributes. Indeed, in $k$-anonymity, the more attributes, the less
homogeneous the records in the $k$-anonymous classes, and the higher
the distortion when generalizing the records in a class or replacing
them by the centroid record. On the other hand, in $\epsilon$-differential
privacy, the privacy budget $\epsilon$ needs to be split among the attributes,
which means that the more attributes, the less budget per attribute
and the more noise needs to be added to each attribute to achieve
 $\epsilon$-differential privacy for the overall data set; hence,
confidentiality increases.
\end{itemize}

\subsection{The permutation model}

In~\cite{Domingomuralidhar16}, we introduced
the permutation model of anonymization.
Consider an original attribute $X=\{x_1, x_2, \cdots, x_n\}$
and its a\-no\-nym\-iz\-ed version $Y=\{y_1, y_2, \cdots, y_n\}$.
Assume $X$ and $Y$  can be ranked (even categorical nominal attributes
can be ranked, using a semantic distance~\cite{rufian}).
For $i=1$ to $n$: compute $j=\mbox{Rank}(y_i)$
and let $z_i=x_{(j)}$, where $x_{(j)}$ is the value of $X$ of rank $j$.
Then call attribute
$Z=\{z_1,z_2, \cdots, z_n\}$ the {\em reverse-mapped} version
of $X$.
For example, if original value $x_1 \in X$ is anonymized
as $y_1 \in Y$, and $y_1$ is, say, the 3rd smallest value in $Y$, then
we take $z_1$ to be the 3rd smallest value in $X$.

\begin{figure*}
\begin{center}
\includegraphics[width=0.7\textwidth]{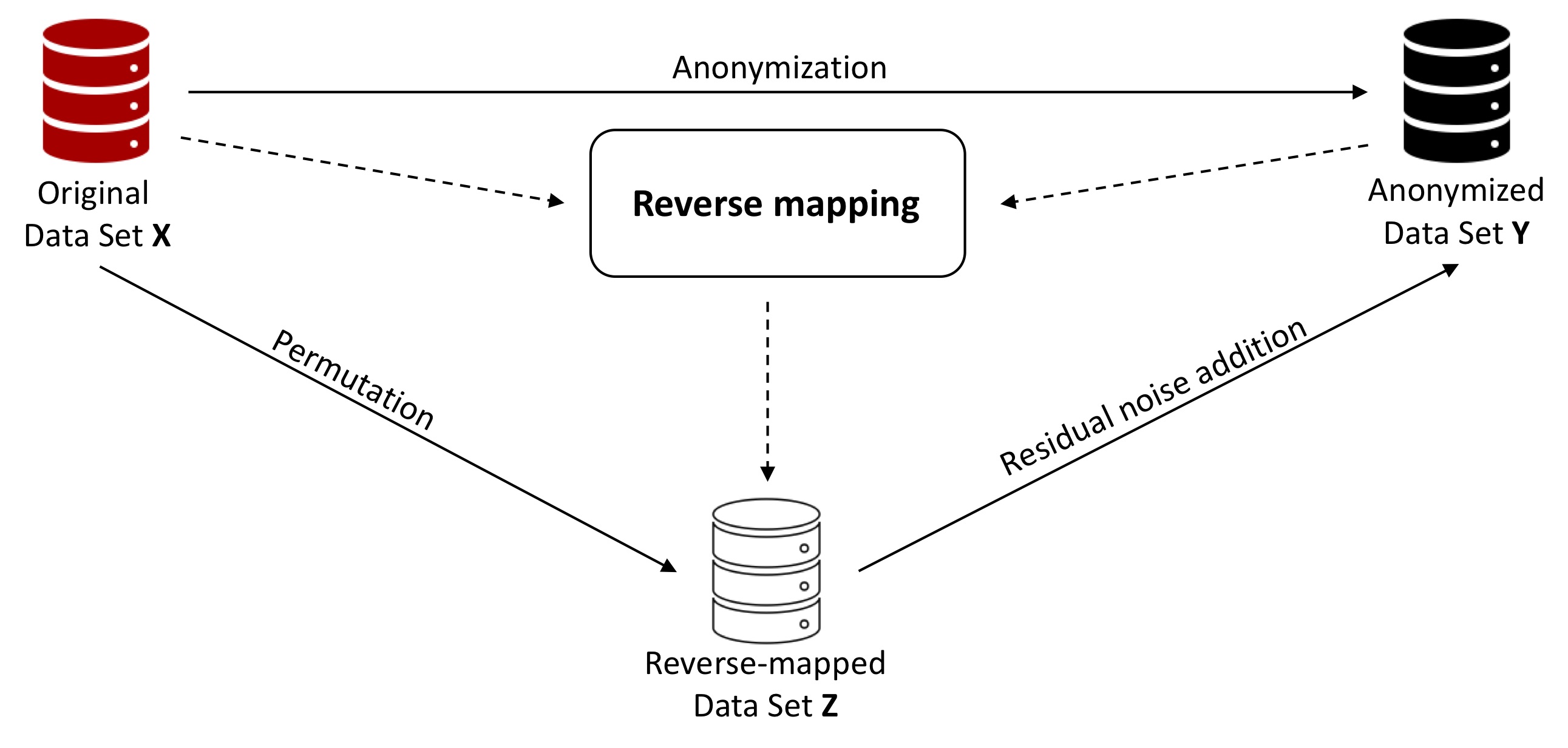}
\end{center}
\caption{The permutation model of anonymization}
\label{permutationmodel}
\end{figure*}

If there are several attributes in the original data set
${\bf X}$ and anonymized data set ${\bf Y}$, the previous reverse-mapping
procedure is conducted for each attribute; call ${\bf Z}$ the data set
formed by reverse-mapped attributes.
Note that: (i) a reverse-mapped attribute $Z$ is a permutation
of the corresponding original attribute $X$; (ii) the rank order
of $Z$ is the same as the rank order of $Y$. Therefore,
{\em any microdata anonymization technique is
 functionally equivalent
to permutation} (from ${\bf X}$ into ${\bf Z}$) {\em followed by residual noise
addition} (from ${\bf Z}$ into ${\bf Y}$). The noise added is residual, because
the ranks of ${\bf Z}$ and ${\bf Y}$ are the same. See
Figure~\ref{permutationmodel}.

\subsection{Canonical correlation}
\label{backcanon}

%JOSEP3. Reviewer 1. Added sentence.
Correlations are range-independent metrics that assess
the relationships between pairs of attributes.
Canonical correlation analysis (CCA) is a multivariate statistics
technique to measure the correlation between two vectors
of random variables~\cite{Hotelling36}. We will use 
CCA to assess the correlation between the original
data set ${\bf X}$, that can be viewed as a sample
of a vector ${\bf x}$ of random variables $X^1,\ldots,X^m$ (the 
original attributes), and the anonymized data set 
${\bf Y}$, that can be viewed as a sample of a vector ${\bf y}$
of random variables $Y^1,\ldots,Y^m$ (the anonymized attributes).

Denote by ${\bf C}_{XX}$ and ${\bf C}_{YY}$ the respective 
covariance matrices of data sets ${\bf X}$ and ${\bf Y}$, 
and by ${\bf C}_{XY}$ the covariance matrix between 
${\bf X}$ and ${\bf Y}$.

The canonical correlations between ${\bf X}$ and ${\bf Y}$ 
can be found by solving the eigenvalue equations
\begin{equation}
\label{canon}
\left\{
\begin{array}{c}
{\bf C}_{XX}^{-1} {\bf C}_{XY} {\bf C}_{YY}^{-1} {\bf C}_{YX} {\bf w}_X = \rho^2 {\bf w}_X \\
{\bf C}_{YY}^{-1} {\bf C}_{YX} {\bf C}_{XX}^{-1} {\bf C}_{XY} {\bf w}_Y = \rho^2 {\bf w}_Y  
\end{array}
\right.
\end{equation}
where the eigenvalues $\rho^2$ are the squared {\em canonical correlations}
and the eigenvectors ${\bf w}_X$ and ${\bf w}_Y$ are 
the normalized canonical correlation {\em basis vectors}.
If both ${\bf X}$ and ${\bf Y}$ have $m$ attributes, there are 
$m$ non-zero solutions of Equations (\ref{canon}), that is,
there are $m$ canonical correlations $\rho_{1},\ldots,\rho_m$,
where we have written them in non-increasing order
$\rho_i \geq \rho_j$ if $i \leq j$.

Only one of the two equations (\ref{canon}) needs to be solved, say
the first one,
because ${\bf w}_X$ and ${\bf w}_Y$ are related as follows:
\[\left\{
\begin{array}{c}
{\bf C}_{XY} {\bf w}_Y = \rho \kappa_X {\bf C}_{XX} {\bf w}_X \\
{\bf C}_{YX} {\bf w}_X = \rho \kappa_Y {\bf C}_{YY} {\bf w}_Y
\end{array}
\right. \]
where
\[ \kappa_X = \kappa^{-1}_Y = \sqrt{\frac{{\bf w}^T_Y {\bf C}_{YY} {\bf w}_Y}{{\bf w}^T_X {\bf C}_{XX} {\bf w}_X}}.\]

Canonical correlation $\rho_1$ turns out to be the correlation
between $u_1={\bf x}^T {\bf w}^1_X$ 
and $v_1= {\bf y}^T {\bf w}^1_Y$, where these linear
combinations of ${\bf x}$ and ${\bf y}$ are the ones yielding
the highest correlation.
Then $\rho_2$ is the correlation between
$u_2={\bf x}^T {\bf w}^2_X$ and $v_2= {\bf y}^T {\bf w}^2_Y$,
where these linear combinations yield the highest 
correlation among the combinations such that $u_1$ and $u_2$ 
are uncorrelated and $v_1$ and $v_2$ are uncorrelated. 
And so on with $\rho_3$ up to $\rho_m$.

See~\cite{Haerdle07} for more details on CCA.
%https://link.springer.com/book/10.1007%2F978-3-662-45171-7

\section{The permutation model and the assessment of confidentiality}
\label{discussion}

\subsection{Confidentiality and the permutation matrices}
\label{intruder}

According to the permutation model, the protection offered 
by an anonymization method comes from two alterations 
of the original data ${\bf X}$: on the one hand, alteration 
of the {\em ranks} of attribute values (that is, permutation
of ${\bf X}$ into ${\bf Z}$)
and, on the other hand, addition of noise (to transform
${\bf Z}$ into ${\bf Y}$) such that it does 
not entail any further change in the ranks.

Hence, the main confidentiality protection principle turns out to be permutation.  
 More precisely,
let us consider the $m$ permutation matrices that respectively represent
the permutation undergone by each of the $m$ attributes. The following
holds:
\begin{itemize}
\item If and only if 
the $m$ permutation matrices are identical, permutation
is trivial in the sense that entire records are swapped,
which provides no confidentiality.
\item If permutation is non-trivial, it provides confidentiality 
as long as the intruder cannot accurately
recreate the $m$ permutation matrices.
\end{itemize}

%JOSEP2. I don't add the sentences "Assessing..." that you suggest, because
%I find it misleading to say that finding the permutation matrics is 
%straightforward if a linkage exists.
The maximum-knowledge intruder 
assumed in the permutation model knows
${\bf X}$ and ${\bf Y}$.
%JOSEP3. Reviewer 2. Added
Thus, this intruder is stronger than any other prior
intruder in the data set anonymization literature.
Furthermore, he is purely malicious: even though 
he already knows the original data set, 
he wants to find the mapping between
the original and the anonymized records,  
 in order to recreate the permutation matrices
and thereby discredit the controller having 
anonymized the data.
 
%JOSEP2. Added this.
In the case of synthetic data, such a natural mapping 
between original and anonymized records does not 
 exist, but the permutation model tells us
that replacing original by synthetic data can still
be viewed as a permutation. 
 A possible
approach is for the intruder to sort 
${\bf X}$ by the $j$-th original attribute 
and ${\bf Y}$ by the $j$-th anonymized attribute, for any 
$1 \leq j \leq m$, and hypothesize that the $i$-th
sorted original record corresponds to the $i$-th anonymized 
record. From that hypothesized mapping, the intruder
may derive hypothesized permutation matrices.

In the remainder of this paper,  ${\bf X}$ 
and ${\bf Y}$ will represent the {\em ranks} of the attributes in the 
original and anonymized data sets, respectively, rather
than their magnitude values. The reason is that our interest 
lies in the permutation of the ranks.

\subsection{Confidentiality and disclosure}

The more accurate the intruder's estimation of the permutation matrices,
the less confidentiality is left and the more chances for disclosure.
%JOSEP3. Moved to the introduction the definitions of the 
%re-identification and attribute disclosure.

Re-identification disclosure cannot be prevented unless
there is a change in ranks, that is, unless ${\bf X} \neq {\bf Z}$
and ${\bf Z}$ is not a trivial permutation of ${\bf X}$.
If  ${\bf X}={\bf Z}$ 
or both data sets are related by a trivial permutation, 
it is immediate for an intruder to link each anonymized
record in ${\bf Y}$ to the record in ${\bf X}$ that has the same
ranks for all attributes. Once the subject's original record
has been determined, re-identification becomes possible.

Let us now look at protection against attribute disclosure.
If ${\bf X}={\bf Z}$ or one data set is a trivial permutation
of the other, 
then protection comes only from noise addition
that transforms ${\bf Z}$ into ${\bf Y}$ but does not change ranks. 
Unless data are very sparse, the noise has 
to be necessarily small, which affords little protection
against attribute disclosure.
Thus, in general, protection against attribute disclosure
necessitates also changes in ranks.

%\section{Covariances to measure permutation-based privacy and utility}
%\label{covariance}

\section{Bounded confidentiality metrics}
\label{privacy}

%JOSEP. Added.
In this section we present three confidentiality metrics.
To compute the first two metrics, one needs to know the 
mapping between records in the original data set 
and records in the anonymized data set. The reason
is that they are based on canonical correlations and 
therefore they require ${\bf C}_{XY}$, the covariance
matrix between ${\bf X}$ and ${\bf Y}$. 
The third metric is based 
on the second metric but it does not need to know
the mapping between records in ${\bf X}$ and ${\bf Y}$.
Thus, it is especially suitable for anonymization via synthetic data.

%JOSEP4. Added
All three metrics use Spearman's rank-based correlation. 
This is a non-parametric (distribution-free) measure 
of the strength
of the monotonic association between two attributes. It  
can be used even when attributes are measured in ordinal scales. 
In certain situations (such as when the relationship between 
the attributes is not linear and/or their distributions are not normal),
 Pearson's product-moment correlation ---more usual and based on attribute
values rather than ranks--- can be unreliable. 
 In these situations, Spearman's 
correlation based on ranks is a better measure than Pearson's~\cite{hauke}.
Furthermore, Spearman's correlation has also been shown to be more
robust than Pearson's in the presence of outliers~\cite{winter} 
and provides higher power for tests of association~\cite{fowler}.

%JOSEP3. Reviewer 1. Added.
Since the proposed metrics are based on the permutation model,
they implicitly assume that the number of anonymized records
is the same as the number of original records. However,
there are SDC methods that may decrease or increase the number 
of records. This is not problematic
as long as the party computing the metrics is the controller
who has performed the anonymization ---which is the usual
situation, because the purpose of the metrics is to guide
anonymization. Indeed, if the number of records 
is reduced due to suppression or sampling, 
the controller knows which 
records have been suppressed and may discard them to obtain 
an original data set with the same number of records as the anonymized
data set. If the number of records decreases or increases 
 as a result of synthetic data generation, then the controller
(or in fact anyone) can sample whichever of the original
or anonymized data sets is larger so that the sampled data set
has the same number of records as the smaller data set.

In the sequel, we assume that the attributes 
in ${\bf X}$ and ${\bf Y}$ are numerical or ordinal categorical,
so that canonical correlations and covariances can be computed
on them. For nominal categorical attributes, an ontology 
can be used whereby a semantically coherent numerical value such as 
marginality~\cite{rufian} can be assigned to each nominal category.
%each nominal category a numerical value derived from its semantic
%distance within a certain ontology, as done in~\cite{rufian}.    
%Then an extension of canonical correlations and covariances ought
%to be designed that allows adapting our confidentiality and 
%utility metrics.
%\end{itemize}

\subsection{Confidentiality metric from the largest canonical correlation}
\label{sec41}

A first approach is to measure confidentiality 
based on the {\em largest} canonical 
correlation $\rho_1$ between the original data set ${\bf X}$ and the 
anonymized data set ${\bf Y}$.

%To focus on the permutation achieved by anonymization, we will 
%consider the {\em ranks} of attributes in ${\bf X}$ and ${\bf Y}$,
%rather than their values.

Since canonical correlations  
are bounded in $[-1,1]$, we can define our permutation-based 
confidentiality metric as
\begin{equation}
\label{privmetric}
CM1({\bf X},{\bf Y}) = 1 - \rho_1^2,
\end{equation}
where, as mentioned above, ${\bf X}$ and ${\bf Y}$ contain ranks 
rather than values. 

According to Expression (\ref{privmetric}),
top confidentiality ($CM1({\bf X},{\bf Y}) = 1$)   
is attained when ranks of attributes in ${\bf Y}$ 
are independent of the ranks of attributes in ${\bf X}$,
which means that the anonymization
can be viewed as a random permutation.

In contrast, zero confidentiality ($CM1({\bf X},{\bf Y}) = 0$) is
achieved            
if the ranks in ${\bf X}$ and ${\bf Y}$ are
the same for {\em at least} one original attribute $X^i$ 
and one anonymized attribute $Y^i$, that is, if the 
anonymization method leaves all ranks unchanged 
for at least
one original attribute.
Ranks can stay unchanged either because the values in 
$X^i$ and $Y^i$ are the same or because
the original values have been perturbed so little that ranks 
are unaffected.
Note that this notion of confidentiality is quite strict: leaving
 a single attribute unprotected brings the confidentiality metric
down to zero. We use the same notion in the next two metrics.

%JOSEP TO KRISH: In Section 5.2 of 
%\url{https://www.cs.cmu.edu/~tom/10701_sp11/slides/CCA_tutorial.pdf}
%they show a connection between canonical correlations
%and Shannon's mutual information $I(X;Y)$.
%It would be good to use mutual information as a privacy metric,
%but the connection they show seems to require that there
%be no higher-order dependencies than correlation and assume
%that variables are Gaussian.
%So I'm not sure we can use it in general.

\subsection{Confidentiality metric from all canonical correlations}
\label{sec42}

A more refined approach is to take all $m$ canonical correlations
into account when measuring confidentiality. 
In~\cite{kay,kay2}, a connection between canonical correlations
and mutual information is shown if the collated 
data sets ${\bf T}=({\bf X},{\bf Y})$, where ${\bf T}$ 
has $2m$ attributes and $n$ records, follow an elliptically
symmetrical distribution (a generalization of the multivariate Gaussian).
The connection is: 
\begin{equation}
\label{mutual}
I(u_i;v_i) = \ln\left(\frac{1}{1-\rho^2_i}\right),
\end{equation}
where $u_i$ and $v_i$ are the linear combinations yielding
$\rho_i$.

Since the pairs $\{(u_i,v_i): i=1, 2,\ldots, m\}$ are mutually
uncorrelated, we can add Expression (\ref{mutual}) for all
pairs to obtain the mutual information between
the original data set ${\bf X}$ and the anonymized data 
set ${\bf Y}$:
\[I({\bf X}; {\bf Y})= \sum_{i=1}^m I(u_i;v_i)\]
\begin{equation}
\label{mutualall}
= \sum_{i=1}^m \ln\left(\frac{1}{1-\rho^2_i}\right)= \ln\left(\frac{1}{\prod_{i=1}^m (1-\rho^2_i)}\right).
\end{equation}

From Expression (\ref{mutualall}) and by analogy with 
Expression (\ref{privmetric}), we can derive the following
confidentiality metric that has the advantages 
of taking all canonical correlations into account
and being related to the mutual information between 
the original and the anonymized data sets.
\begin{equation}
\label{privmetricall}
CM2({\bf X},{\bf Y}) = \prod_{i=1}^m (1-\rho^2_i) \left[ = e^{-I({\bf X};{\bf Y})}\right].
\end{equation}

The second equality between brackets in Expression (\ref{privmetricall}) 
can only be guaranteed if the above distributional assumptions hold,
in which case
Expression (\ref{privmetricall}) 
 can be justified 
using mutual information.

Regardless of the distributional assumptions, 
$CM2({\bf X},{\bf Y})$ can be computed from the 
canonical correlations and 
the following holds:
\begin{itemize}
\item Top confidentiality $CM2({\bf X},{\bf Y})=1$
is reached when
the anonymized data set and the original
data sets tell nothing about each other, which 
is the same as saying that mutual information between
them is $I({\bf X};{\bf Y})=0$.
\item Zero confidentiality $CM2({\bf X},{\bf Y})=0$ occurs if at least one
of the canonical correlations is 1. This occurs 
if at least one original attribute is disclosed 
when releasing ${\bf Y}$. Since $\rho_1$ is the largest
correlation, this means that we have $CM2({\bf X},{\bf Y})=0$
if and only if $\rho_1=1$,
 in which case we also have 
that the metric of Expression (\ref{privmetric}) is $CM1({\bf X},{\bf Y})=0$.
\end{itemize}

\subsection{Mapping-free confidentiality metric} 
\label{sec43}

The confidentiality metrics defined in Sections~\ref{sec41} and~\ref{sec42}
implicitly assume a known mapping between records in the original
data set ${\bf X}$ and records in the anonymized data set 
${\bf Y}$ to compute canonical correlations (in particular
to compute the covariance ${\bf C}_{XY}$ between ${\bf X}$ and ${\bf Y}$). 
Such a mapping is naturally known 
to the controller if she obtains 
each record in ${\bf Y}$ by masking 
a record in ${\bf X}$, via noise addition or another SDC method.

However, in the case of synthetic data generation there is no 
natural mapping 
between original and anonymized values. Indeed, data synthesis
generates a complete data set by using the distributional 
characteristics of the original data set, rather than the original data
themselves. Many synthetic data generation procedures
have been proposed in the 
literature~\cite{burridge,shuffling,reiter,drechsler}. Since the {\em individual} records 
in the synthetic data set ${\bf Y}$ do not depend on the 
{\em individual} records in the original data set ${\bf X}$, the 
correlation between original data and synthetic data 
can be expected to be zero, subject to the sampling error.
Consequently, metrics $CM1$ and $CM2$ will always be close to 1, 
even if the synthetic data leak the original data (see
examples further down in this section and in Section~\ref{empirical}).

%The above poses a serious problem, because even trivial 
%synthetic data generation procedures may seem to achieve
%high data confidentiality. 
In this section, we propose a confidentiality metric that
does not need to know in advance the mapping between 
records of ${\bf X}$ and ${\bf Y}$.
It uses the permutation model and more specifically reverse 
mapping~\cite{Domingomuralidhar16},
whereby the values of an anonymized attribute can be viewed
as a permutation of the values of the corresponding original attribute
(plus perhaps a marginal amount of noise).
Hence, even if anonymized values look uncorrelated with the 
original values, a permutation linking anonymized and original
values exists. As mentioned in 
Section~\ref{intruder}, a maximum-knowledge intruder knowing ${\bf X}$
and ${\bf Y}$ can try to guess  
the mapping between records across both data sets 
by sorting ${\bf X}$ and ${\bf Y}$  by one attribute and evaluating
%JOSEP3. Reviewer 2. Corrected verb placement.
how similar the values of the rest of attributes are 
in the sorted data sets.

To reflect the above procedure, we propose the 
confidentiality metric $CM3$ computed by Algorithm~\ref{algCM3}.
\vspace{1ex}\\

\begin{algorisme}\label{algCM3}~\\
\begin{enumerate}
\item For $j=1$ to $m$ do:
\begin{enumerate}
\item Sort the original data set by its $j$-th attribute 
and let ${\bf X}^{-j}$ be the projection of the 
sorted data set on all attributes except the $j$-th one.
\item Sort the anonymized data set by its $j$-th attribute
and let ${\bf Y}^{-j}$ be the projection of the 
sorted data set on all attributes except the $j$-th one.
\item Compute $CM2({\bf X}^{-j},{\bf Y}^{-j})$
according to Expression (\ref{privmetricall}).
\end{enumerate}
\item Let 
\begin{equation}
\label{CM3}
CM3({\bf X}, {\bf Y}) = \min_{1\leq j \leq m} CM2({\bf X}^{-j},{\bf Y}^{-j}).
\end{equation}
\end{enumerate}
\end{algorisme}

The $CM3$ confidentiality metric can be readily applied when
${\bf Y}$ is synthetic: a mapping between records
in ${\bf X}$ and ${\bf Y}$ is not needed because 
one tries all $m$ possible mappings obtained when using each single
attribute as a sorting key.  

The following are interesting cases
of synthetic data sets:
\begin{itemize}
\item Let ${\bf X}$ be such that attributes $X^i$ and $X^j$
are perfectly correlated. Assume that the synthetic ${\bf Y}$ 
also preserves the relationship between $Y^i$ and $Y^j$
to be the same as the one between $X^i$ and $X^j$. 
In other words, the permutations from $X^i$ 
to $Y^i$ and from $X^j$ to $Y^j$ are
exactly the same. Hence, if we sort ${\bf X}$ by $X^i$
and ${\bf Y}$ by $Y^i$, attributes 
$X^j$ in ${\bf X}^{-i}$ and $Y^j$ in ${\bf Y}^{-i}$ 
are perfectly correlated.
Thus $CM2({\bf X}^{-i},{\bf Y}^{-i})=0$ and in consequence 
$CM3({\bf X},{\bf Y})=0$.
However directly using $CM2$ on ${\bf X}$ and ${\bf Y}$
yields in general $CM2({\bf X},{\bf Y})\neq 0$.
\item If the attributes in ${\bf X}$ are very highly
correlated, any masking method that preserves the correlation 
structure of ${\bf X}$ in ${\bf Y}$ cannot permute much. 
Consequently, it offers less confidentiality than if 
the correlation structure was not preserved. 
Equation (\ref{CM3}) captures this situation of rank 
preservation among ${\bf X}$ and ${\bf Y}$ and gives
a low value for $CM3$, even if $CM1$ and $CM2$ may be quite
high, as illustrated in an experiment in Section~\ref{empiricalIPSO}.
\end{itemize}

%JOSEP3. Reviewer1. 
\subsection{Summary on confidentiality metrics}

$CM1$ or $CM2$ should be applied whenever
the mapping between original records and anonymized records
is known. 
%While $CM1$ is simpler to compute, $CM2$ is more
%refined and captures more information.
If the mapping is not known, 
such as in synthetic data, then $CM3$ should be applied.

The following holds regarding $CM1$ and $CM2$:
\begin{itemize}
\item $CM2$ is a product of terms not greater than 1 whose
first term is $CM1$. Hence $CM2$ is not greater than $CM1$.
\item $CM2=0$ if and only if $CM1=0$, because $\rho_1$ is the 
largest correlation.
\item Since $CM2$ takes all canonical correlations into account,
it is a better metric than $CM1$, although $CM1$ is easier to compute.
\end{itemize}

It is difficult to compare $CM3$ with $CM1$ or $CM2$. 
The former is intended for use with synthetic microdata 
in which there is no linkage between the records in the original and 
masked data. When applied to non-synthetic masking methods, 
since $CM3$ is based on an arbitrary linkage and $CM2$ 
is based on the true linkage, we normally have $CM3 > CM2$. 
The exception is the case of trivial 
permutation (swapping entire records), 
in which $CM3=0$.
 Indeed, under trivial permutation
 sorting
${\bf X}$ by any attribute $X^i$ and ${\bf Y}$ by
the corresponding $Y^i$ yields identical sorted
data sets, and thus $CM3=0$.

Therefore, $CM3$ has the advantage 
of detecting trivial permutation, which $CM1$ and $CM2$
do not detect. 

$CM3$ can also be viewed as the confidentiality metric 
from the intruder's perspective. Unlike the data controller, 
the intruder does not know the true linkage between original 
and masked records and may evaluate confidentiality using $CM3$. 
Interestingly, when synthetic microdata are released neither the 
data controller nor the intruder know the ``true'' linkage and 
$CM3$ is a natural confidentiality metric for both. 

\section{A covariance-based bounded utility metric}
\label{utility}

In statistical disclosure control, a 
confidentiality metric needs to have a companion utility metric 
to allow for the necessary trade-off evaluation between utility
and confidentiality.
Since in Expressions (\ref{privmetric}), (\ref{privmetricall})
and (\ref{CM3})
 we have proposed confidentiality metrics that are bounded between
0 and 1, we need companion utility metrics that are also bounded.

%JOSEP3. Revisor 1. Expanded/modified paragraph.
After measuring confidentiality in terms of covariance
matrices, it is natural to
examine whether covariances can also conveniently 
characterize utility. 
Although some utility metrics focus on the mean
error between original and anonymized data, 
 preserving the covariance structure
 seems the most relevant utility feature
for all those analyses aimed at discovering relationships
between attributes.

If the attributes in the original and the anonymized data set 
are Gaussian (resp. near-Gaussian), then they are fully 
(resp. almost fully) described by their second-order statistics. 
Hence, in this case the covariance matrix is a sufficient 
measure of utility.
The more the distribution of the attributes departs from Gaussian,
%JOSEP3. Reviewer 2. Verb placement fixed.
the more likely higher-order relationships are 
 that stay uncaptured by the covariance matrix, which nonetheless
remains a meaningful utility measure.

Let a data set ${\bf X}$ be masked as ${\bf Y}$.
As said above, we will consider the ranks of values in both data sets,
rather than the values themselves.
If all attributes are numerical and sparse, one might choose 
to work on values rather than ranks in order to capture
utility more closely.

%\subsection{Utility based on eigenvalues 
%and eigenvectors of covariance matrices}

In terms of covariances, maximum utility occurs when
${\bf C}_{XX}={\bf C}_{YY}$, in which case the 
(second-order) relationships between
attributes in the original data set are exactly preserved
in the masked data set.
%JOSEP3. Reviewer 2. Verb placement fixed.
To compare how similar ${\bf C}_{XX}$ and  ${\bf C}_{YY}$ are, 
a rough procedure is 
to compare their respective eigenvalues.
Let the eigenvalues of ${\bf C}_{XX}$ be $\lambda^X_1$, $\ldots$,
$\lambda^X_m$, and the eigenvalues of ${\bf C}_{YY}$ be 
$\lambda^Y_1$, $\ldots$, $\lambda^Y_m$.
In the case of a covariance matrix, the first eigenvalue
represents the magnitude of the maximum spread 
of the data, the second eigenvalue is the magnitude
of the second largest data spread in a direction
orthogonal to the maximum spread direction, etc.
Thus, eigenvalues appear in non-increasing order.
The case of all $m$ eigenvalues of a covariance matrix 
being equal would reflect a set of 
records having equal spread in all directions of the 
$m$-dimensional space, 
a sort of $m$-dimensional sphere; this occurs when all
attributes are uncorrelated.

%To obtain a bounded metric
%comparing eigenvalues, we scale eigenvalues so that they lie 
%in the $[0,1]$ interval and add to 1.
%Let $\hat{\lambda}^X_1$, $\ldots$,
%$\hat{\lambda}^X_m$, and
%$\hat{\lambda}^Y_1$, $\ldots$, $\hat{\lambda}^Y_m$ be 
%the scaled eigenvalues in non-increasing order. 
%We can then compare scaled eigenvalues as a difference
%of Shannon entropies:
%\begin{equation}
%\label{utilmetric}
%|\sum_{j=1}^m \hat{\lambda}^X_j \log_2 \frac{1}{ \hat{\lambda}^X_j} -
%\sum_{j=1}^m \hat{\lambda}^Y_j \log_2 \frac{1}{ \hat{\lambda}^Y_j}|.
%\end{equation}

%If the scaled eigenvalues of covariance matrices in 
%both data sets are the same, 
%Expression (\ref{utilmetric}) yields 0.
%On the other hand, 
%Shannon's entropy in the case of $m$ values is 
%bounded between 0 and $\log_2 m$. Thus,
%Expression (\ref{utilmetric}) reaches
%a maximum $\log_2 m$ when one data set has
% entropy 0 for the scaled eigenvalues of its covariance matrix
%and the other has maximum entropy $\log_2 m$.
%This corresponds to one data set being a straight line 
%in the $m$-dimensional space
%(in which case the first scaled eigenvalue is 1 and the rest 
%0) and the other data set being an $m$-dimensional sphere
%(in which case all scaled eigenvalues are $1/m$).
%The first case indicates perfectly correlated attributes and
%the second uncorrelated attributes.

Unfortunately, just comparing eigenvalues
is not sufficient to assess utility, 
because eigenvalues capture only the magnitude of the maximum spreads
on orthogonal directions, but not the directions themselves.
Thus, two data sets can share the same set of eigenvalues
while being different: in particular, if ${\bf Y}$ is a rotation
of ${\bf X}$, both data sets have the same eigenvalues.

For a given spread magnitude (eigenvalue), the direction of spread
is described by the corresponding eigenvector.
We loosely adapt a procedure proposed in~\cite{Garc12} for comparing 
covariance matrices.
Let $\lambda^X_1$, $\ldots$,
$\lambda^X_m$, resp.
$\lambda^Y_1$, $\ldots$, $\lambda^Y_m$ be 
the eigenvalues of ${\bf C}_{XX}$, resp. ${\bf C}_{YY}$ 
in non-increasing order. 
Let ${\bf v}^X_1$, $\ldots$, ${\bf v}^X_m$, resp.
${\bf v}^Y_1$, $\ldots$, ${\bf v}^Y_m$ be the 
corresponding eigenvectors of ${\bf C}_{XX}$, resp. ${\bf C}_{YY}$.
Then it holds that
\[ \lambda^X_j = ({\bf v}^X_j)^T {\bf C}_{XX} {\bf v}^X_j, \;\; j=1,\ldots, m.\]
Now consider
\[ \lambda^{Y|X}_j = ({\bf v}^X_j)^T {\bf C}_{YY} {\bf v}^X_j, \;\; j=1,\ldots, m.\] 
Just as each eigenvalue $\lambda^X_j$ can be viewed as the proportion
of the variance of the attributes in ${\bf X}$ explained by 
the corresponding eigenvector ${\bf v}^X_j$, we can view
$\lambda^{Y|X}_j$ as the proportion of the variance of 
the attributes in ${\bf Y}$ explained by  ${\bf v}^X_j$. 
Note that the values $\lambda^{Y|X}_j$, for $j=1, \ldots, m$, are not necessarily 
non-increasing.

The highest level of utility occurs when 
$\lambda^X_j = \lambda^{Y|X}_j$ for $j=1,\ldots, m$,
which occurs when ${\bf C}_{XX}={\bf C}_{YY}$. 

Covariance matrices are positive semi-definite, which means
that their eigenvalues are all non-negative.
If $\lambda_1, \ldots, \lambda_m$ are the eigenvalues of a covariance
matrix, let $\hat{\lambda}_1, \ldots, \hat{\lambda}_m$ be
their scaled versions so that they add to 1. 
Then the extent to which ${\bf C}_{XX}$ and ${\bf C}_{YY}$ differ
can be expressed as 
\begin{equation}
\label{diffcovariance}
\sum_{j=1}^m (\hat{\lambda}^X_j - \hat{\lambda}^{Y|X}_j)^2. 
\end{equation}

\begin{proposition}
\label{prop1}
Expression (\ref{diffcovariance}) is bounded between 0 and 2.
The maximum occurs when all the variance of ${\bf X}$ occurs
in a single direction and all the variance of ${\bf Y}$ also occurs
in a single direction that is orthogonal to the previous one.
\end{proposition}

{\em Proof:} The minimum is clearly 0. To compute the 
maximum, take into account that
$\sum_{j=1}^m \hat{\lambda}^X_j = \sum_{j=1}^m \hat{\lambda}^{Y|X}_j =1$.
Thus, the maximum occurs when two of the squares added in 
Expression (\ref{diffcovariance}) are 1, and a square
can be 1 if it is either $(1-0)^2$ or $(0-1)^2$.
This situation occurs when:
i) all the variance of ${\bf X}$ 
is explained by the first eigenvector ${\bf v}^X_j$, in which
case we have $\hat{\lambda}^X_1 =1$ and 
$\hat{\lambda}^X_l = 0$ for all $l=2,\ldots,m$; and ii) all 
the variance of ${\bf Y}$ is explained by one eigenvector
${\bf v}^X_{j'}$ with $j' \neq 1$ and hence orthogonal to ${\bf v}^X_j$,
in which case $\hat{\lambda}^{Y|X}_{j'} =1$ and the rest 
of $\hat{\lambda}^{Y|X}_j$ are zero. \hfill $\Box$
%Consequently, Expression (\ref{diffcovariance}) is $(1-0)^2+(0-1)^2=2$. $\Box$

According to Proposition~\ref{prop1}, the maximum difference
between two covariance matrices ${\bf C}_{XX}$ and ${\bf C}_{YY}$ 
can be quantified as 2. However,
value 2 is reached when ${\bf C}_{YY}$ has a very specific shape
with respect to ${\bf C}_{XX}$. Rather, we are interested in 
finding a measure of utility, that is, to see how much
the covariances in ${\bf X}$ are preserved in ${\bf Y}$. In this sense,
the intuition is that the maximum utility loss occurs when 
the covariances of ${\bf X}$ are completely lost in ${\bf Y}$, 
or equivalently when
%JOSEP2. Corrected because no variance would be \hat{\lambda}^{Y|X}_j=0
any of the $m$ eigenvectors of ${\bf C}_{XX}$ 
explains a fraction $1/m$ of the variance of ${\bf Y}$. 
In this case, 
$\hat{\lambda}^{Y|X}_j=1/m$ for $j=1,\ldots, m$, and 
Expression (\ref{diffcovariance}) becomes:
\begin{equation}
\label{diffcovariance2}
\sum_{j=1}^m (\hat{\lambda}^X_j - 1/m)^2. 
\end{equation}

\begin{proposition}
\label{prop2}
Expression (\ref{diffcovariance2}) is bounded between 0 and $(m-1)/m$.
\end{proposition}

{\em Proof:} The minimum is clearly 0. The maximum is reached 
when all the variance of ${\bf X}$ occurs
in a single direction. In that situation:
\[\sum_{j=1}^m (\hat{\lambda}^X_j - 1/m)^2 = (1-1/m)^2 + (0-1/m)^2 + \ldots + (0-1/m)^2\]
\[= (m-1)/m. \] \hfill $\Box$

We are now in a position to define the following 
utility measure based on Expressions (\ref{diffcovariance}) 
and (\ref{diffcovariance2}):
\begin{equation}
\label{utilitymeasure}
UM({\bf X}, {\bf Y}) = \left\{ 
\begin{array}{l}
1 \mbox{~~~if $\hat{\lambda}^X_j = \hat{\lambda}^{Y|X}_j=1/m$ for $j=1,\ldots, m$;} \\
1 - \min\left(1, 
\frac{\sum_{j=1}^m (\hat{\lambda}^X_j - \hat{\lambda}^{Y|X}_j)^2}{\sum_{j=1}^m (\hat{\lambda}^X_j - 1/m)^2}\right) \mbox{~~~otherwise.}
\end{array}
\right.
\end{equation}
The first case in Expression (\ref{utilitymeasure})
covers the (very exceptional) situation in which both the 
original data set and the anonymized data set are perfectly 
uncorrelated, which means there is no utility loss.
Regarding the second case,
from Propositions~\ref{prop1} and~\ref{prop2}, 
the ratio within the argument of the minimum function can be greater than 1.
By using the minimum, we make
sure $UM({\bf X}, {\bf Y})$ is bounded between 0 and 1. Thus we have:
\begin{itemize}
\item Top utility $UM({\bf X}, {\bf Y})=1$ is reached when information 
loss is zero, which occurs when $\hat{\lambda}^X_j = \hat{\lambda}^{Y|X}_j$ 
for $j=1,\ldots, m$.
\item Zero utility $UM({\bf X}, {\bf Y})=0$ occurs if 
 $\hat{\lambda}^X_j$ and $\hat{\lambda}^{Y|X}_j$ differ at least
as much as $\hat{\lambda}^X_j$ and the eigenvalues
of an uncorrelated data set. 
\end{itemize}

\section{Empirical work}
\label{empirical}

The purpose of the experiments reported in this section
is to highlight that the confidentiality and utility metrics
presented above can be applied to a variety of privacy-first
and utility-first anonymization
approaches.
Specifically, we consider privacy models 
($k$-anonymity~\cite{Samarati98} and 
differential privacy~\cite{Dwork06})
and SDC methods (additive noise,
multiplicative noise and synthetic data).

\subsection{Privacy models}

To test $k$-anonymity and $\epsilon$-differential privacy, 
we took as original data set the
``Census'' data set,
which contains 1,080 records with numerical attributes~\cite{Brand}.
This data set was used in the European project CASC and
in~\cite{Domi01creta,Dand02b,Yanc02,Lasz05,Domingo05,Domi08cam,Domingo10,Domingo14}.
Like in~\cite{Domingo10,Domingo14}, we took attributes
FICA (Social security retirement payroll deduction),
FEDTAX (Federal income tax liability),
INTVAL (Amount of interest income) and POTHVAL (Total other persons income).
We considered all four attributes as quasi-identifiers in all
of our tests.
The resulting records were all different from each other.
Since all attributes represent
non-negative amounts of money, we took as boundaries for the 
domain of each attribute $0$ and $1.5$ times 
the maximum value of the attribute
in the data set.

%JOSEP3. Changed to reflect new experiments.
We then took three versions of the ``Census'' data set: 
one with all 4 attributes, one with 3 attributes (FICA, FEDTAX and INTVAL) 
and one with 2 
attributes (FICA and FEDTAX). 
We separately anonymized the three versions as follows:
\begin{itemize}
\item Achieving
$k$-anonymity for $k=2,3,\ldots,100$ and 
$k=200, 300, 400, 500$ using the MDAV
microaggregation algorithm~\cite{Domingo05}.
\item Achieving $\epsilon$-differential privacy
via Laplace noise addition to unagreggated attribute
 data for $\epsilon$= {0.01, 0.1, 1, 10, 25, 50, 100}, which
covers the usual range of
differential privacy levels observed in the
literature~\cite{Dwork11,Char10,Char12,Mach08} plus
some very large $\epsilon$ values. For 
each $\epsilon$ value, five differentially private data sets were 
generated and 
 utility and confidentiality metrics were averaged over the 
five data sets.
\end{itemize}

\begin{figure*}
\begin{center}
\includegraphics[width=0.45\textwidth]{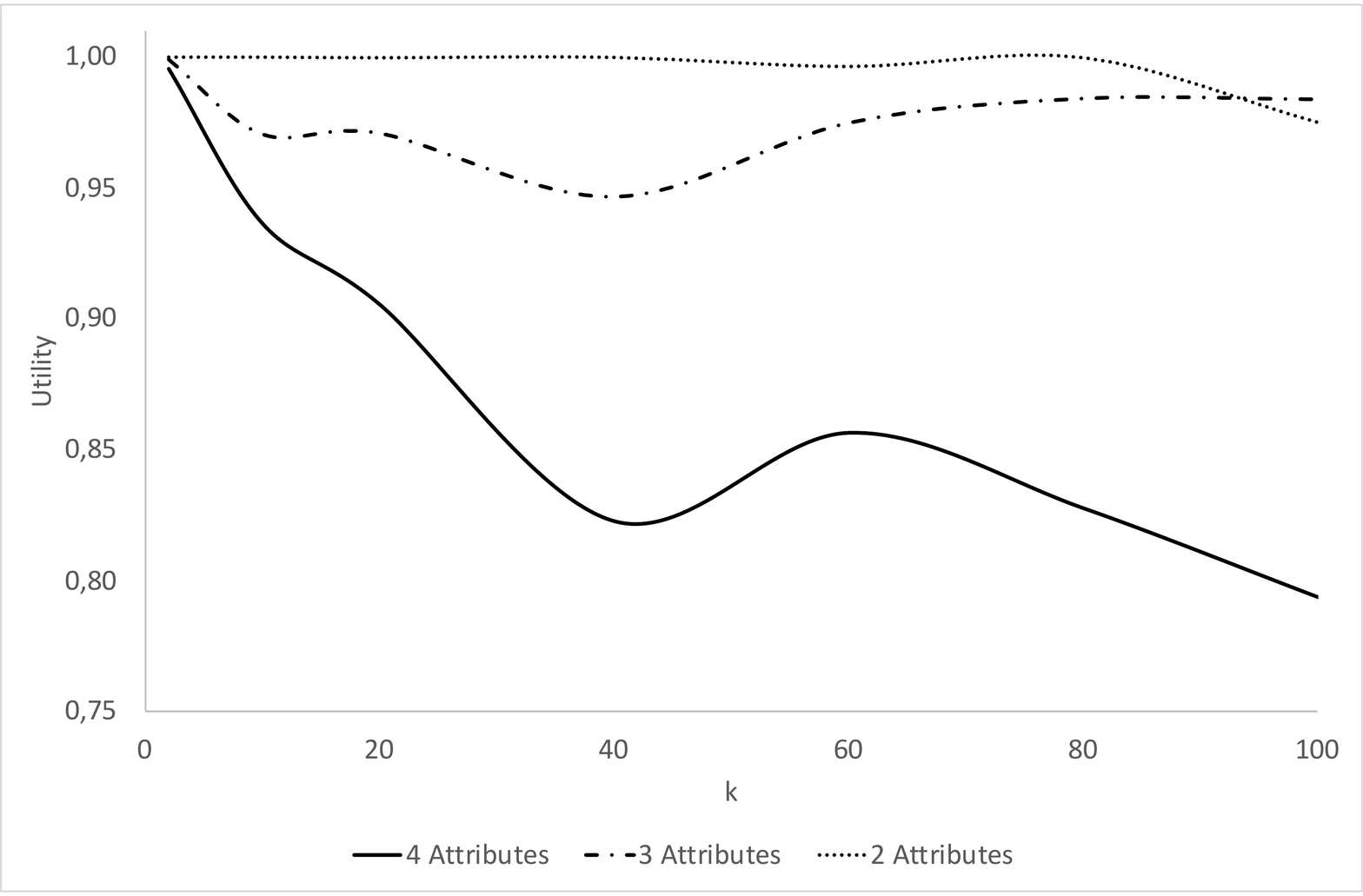}\\
\includegraphics[width=0.45\textwidth]{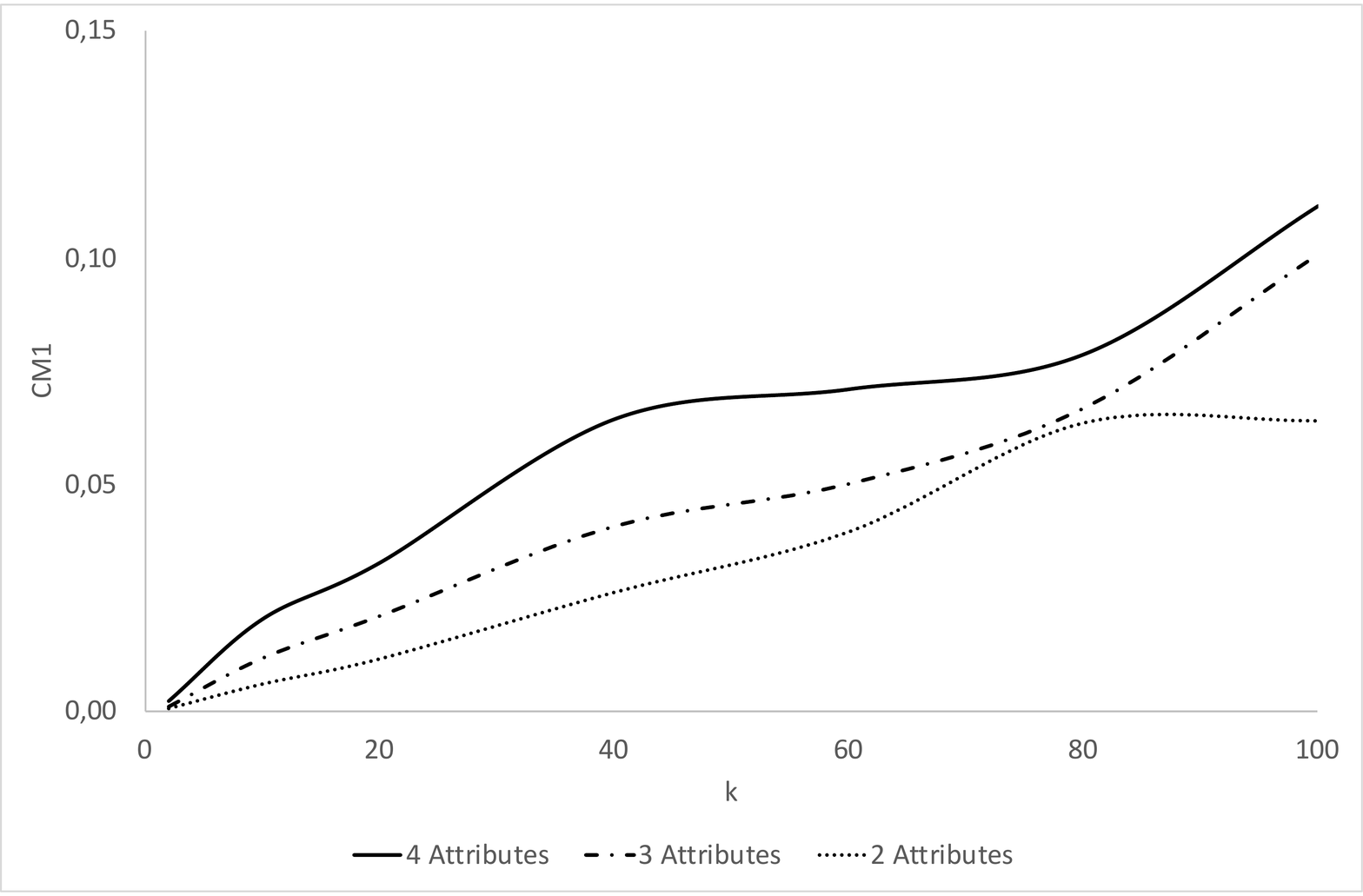}
\includegraphics[width=0.45\textwidth]{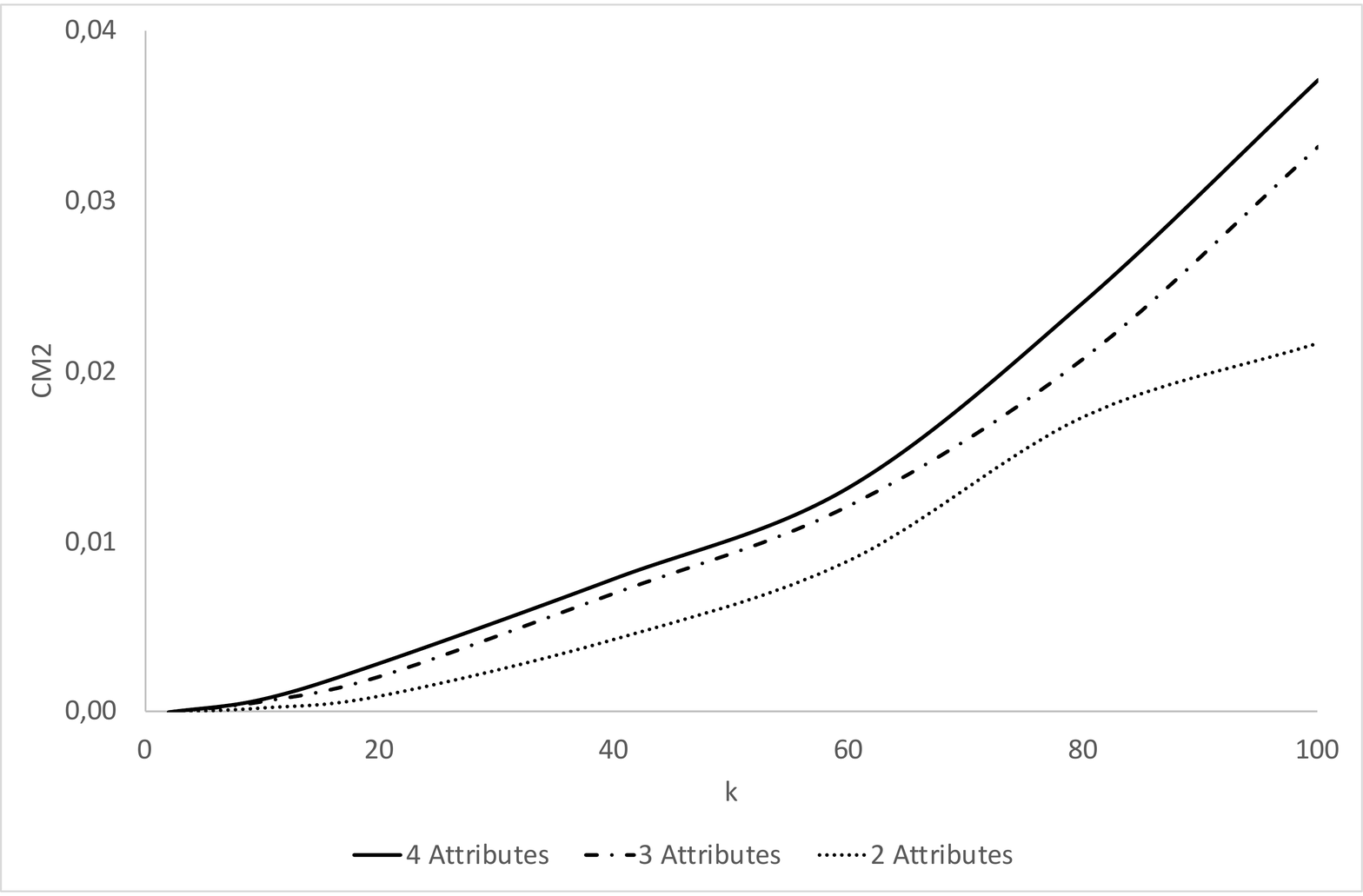}
\end{center}
%JOSEP3. Added number of attributes in caption.
\caption{Utility metric (top) and
confidentiality metrics $CM1$ (bottom left) and $CM2$ (bottom right) 
for $k$-anonymized data
as a function of $k$ and the number of attributes}
\label{kanom}
\end{figure*}

Figure~\ref{kanom} shows the utility
metric and the confidentiality metrics $CM1$ and $CM2$ for the $k$-anonymized data as a function of $k$
%JOSEP3. Added.
and the number of attributes.
As expected, as $k$ increases, utility decreases
and confidentiality increases. 
%JOSEP3. Added.
Also, as anticipated in Section~\ref{privconf}, for fixed $k$
a decrease of utility and an increase of confidentiality 
is observed when the number of attributes grows.
  
\begin{figure*}
\begin{center}
\includegraphics[width=0.45\textwidth]{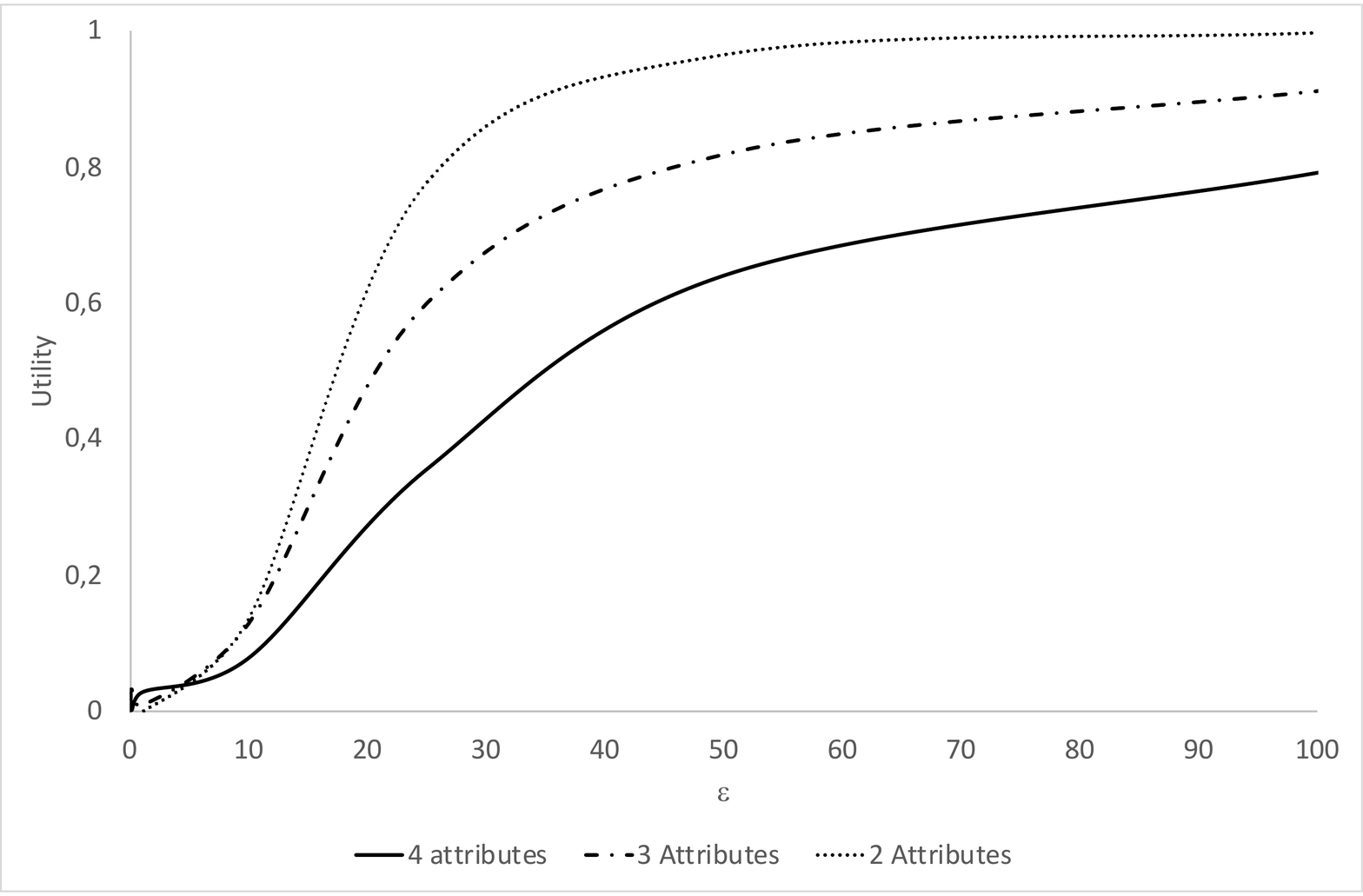}\\
\includegraphics[width=0.45\textwidth]{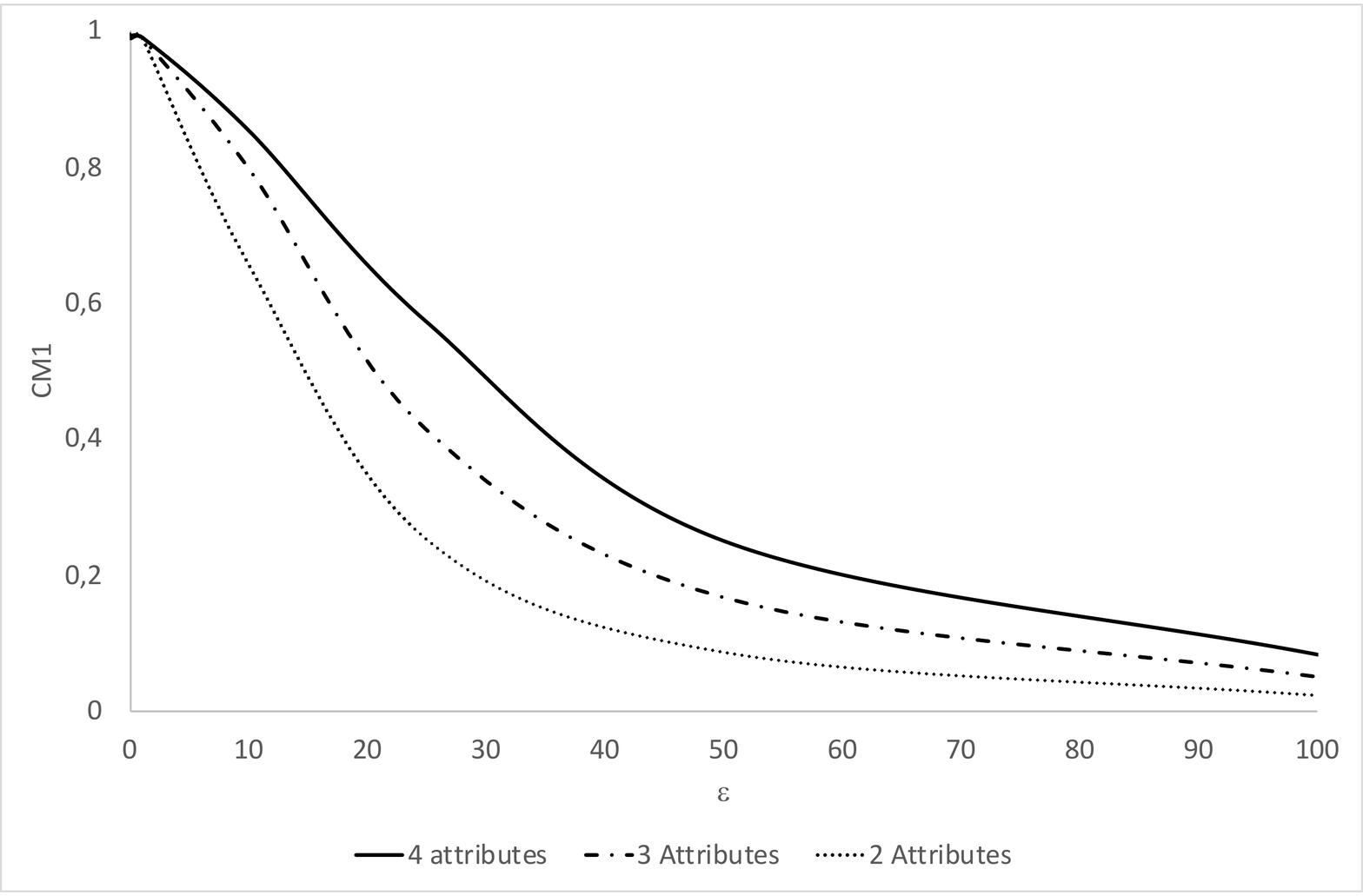}
\includegraphics[width=0.45\textwidth]{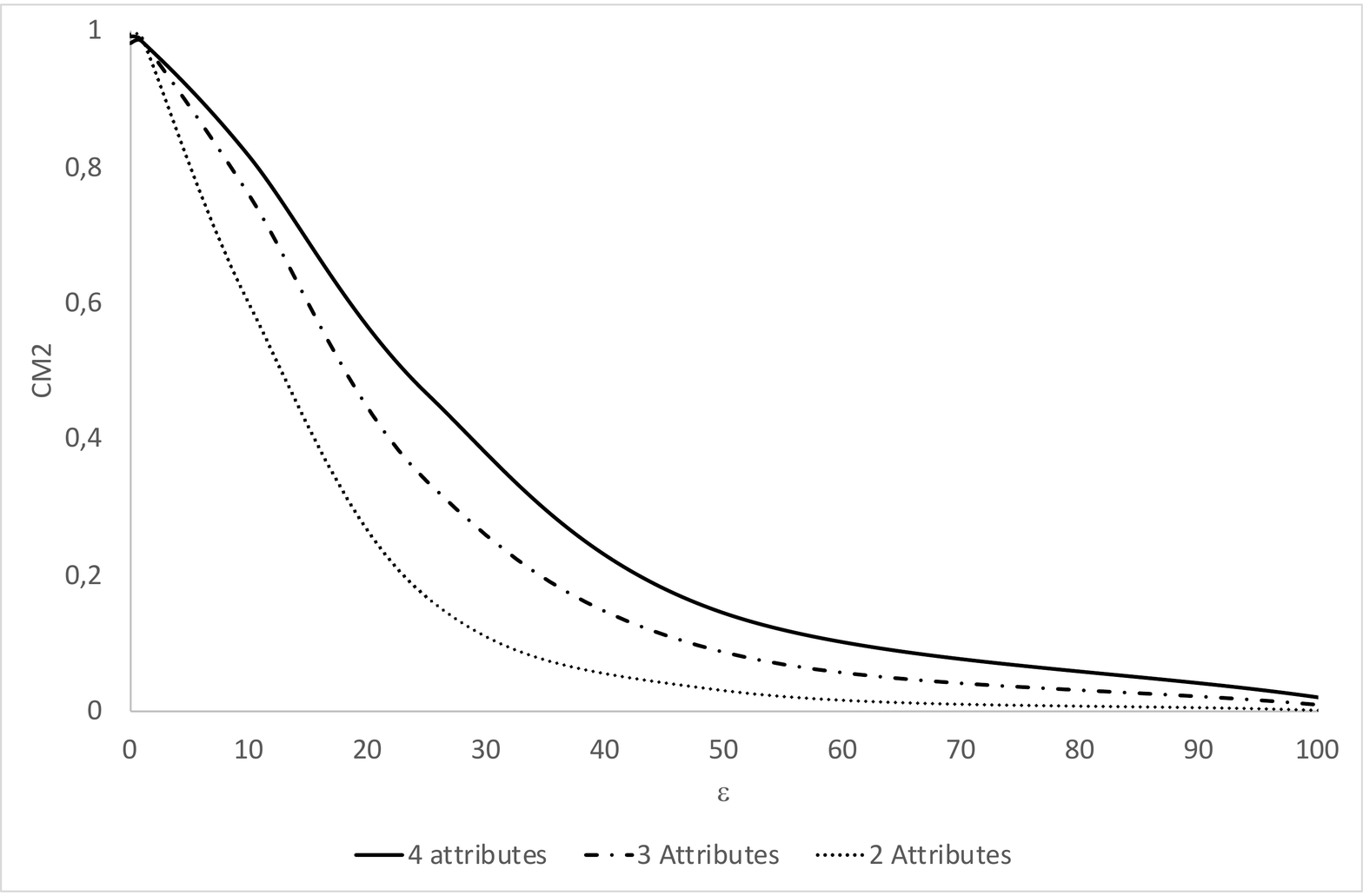}
\end{center}
%JOSEP3. Added number of attributes in caption.
\caption{Utility metric (top) and confidentiality metrics $CM1$ 
(bottom left) and $CM2$ (bottom right)
for $\epsilon$-differentially
private data as a function of $\epsilon$ and the number of attributes}
\label{dp}
\end{figure*}

Figure~\ref{dp} shows the utility metric and the confidentiality
metrics $CM1$ and $CM2$ for 
the $\epsilon$-differentially private data as a function of
$\epsilon$ 
%JOSEP3. Added.
and the number of attributes.
As expected, as $\epsilon$ increases, utility increases
and confidentiality decreases.
%JOSEP3. Added.
Also, consistently with Section~\ref{privconf},
for fixed $\epsilon$
a decrease of utility and an increase of confidentiality
is observed when the number of attributes 
%JOSEP11. Added Krish's footnote here.
grows\footnote{ 
The particular data set that we used in this study consists of skewed 
economic data. The range and hence the global sensitivity of attributes
in the data set are large.   
As a result, the variance of the noise added is also relatively 
large even when $\epsilon=100$. In addition, with multiple attributes,
the $\epsilon$ budget must be split among the attributes, effectively 
reducing the budget for each attribute. 
For four attributes, the variance of the Laplace noise added is four times 
the variance of the Laplace noise added for two attributes. Consequently, 
the correlation between the original and masked data in the four-attribute
case is substantially lower than in the two-attribute case.
Furthermore, canonical correlation evaluates correlation among {\em all} 
original and masked attributes simultaneously. Not only is the variance added in the case of four attributes four times larger than in the case of two attributes, but in the former case four masked attributes are compared against four original attributes, which also contributes to a greater utility loss than comparing two masked attributes against two original attributes as in the latter case. 
If utility is an important consideration, 
the data controller may wish to consider an alternative mechanism 
for implementing differential privacy when there are many attributes.}. 

%So, {\em our metrics are consistent
%with the intuition} behind the two main privacy models in the literature.

Note that, since $k$-anonymity is achieved via microaggregation
and $\epsilon$-differential privacy via noise addition, for 
both privacy models the controller knows the correspondence 
between each anonymized record and the original 
record it derives from. Therefore, it does not make sense
for the controller to use the mapping-free confidentiality metric
$CM3$.  

By superposing Figures~\ref{kanom} and~\ref{dp} 
(or rather the numbers behind them), one can 
compare the utility-confidentiality trade-offs 
 achieved by $k$-anonymity and $\epsilon$-differential privacy. 
For most parameters we tried on the three versions 
of the ``Census'' data set, $k$-anonymity yields
substantially more utility (above $0.9$ for all $k$) 
but substantially less confidentiality
than differential privacy. 

%\item $k=100$ provides slightly more utility ---$UM=0.793$ vs $UM=0.789$--- and
%confidentiality than $\epsilon=25$ 
%---$(CM1,CM2)=(0.112,0.037)$ vs $(CM1,CM2)=(0.084,0.022)$;
%\item $k=300$ provides roughly the same confidentiality
%---$(CM1,CM2)=(0.360,0.219)$ vs $(CM1,CM2)=(0.340,0.234)$--- as 
%$\epsilon=10$, but substantially more utility
%---$UM=0.753$ vs $UM=0.579$.
%\end{itemize}

With privacy models, the data controller is blind regarding
 any issue other than privacy. In addition, it is also 
very difficult to compare
across privacy models. One of the objectives of this paper is
 to propose measures that inform the controller on other
 aspects of the masking procedure, namely utility and confidentiality.
 This allows the controller to compare
the trade-offs offered by different privacy models 
and to evaluate whether a lower/higher level
 of privacy may be warranted for the data set. The decision regarding
the right levels of anonymity, confidentiality and utility must be 
made by the data controller. 
 Our measures give him information to make that decision. 

\subsection{Noise-based SDC methods}

\begin{figure*}
\begin{center}
\includegraphics[width=0.7\textwidth]{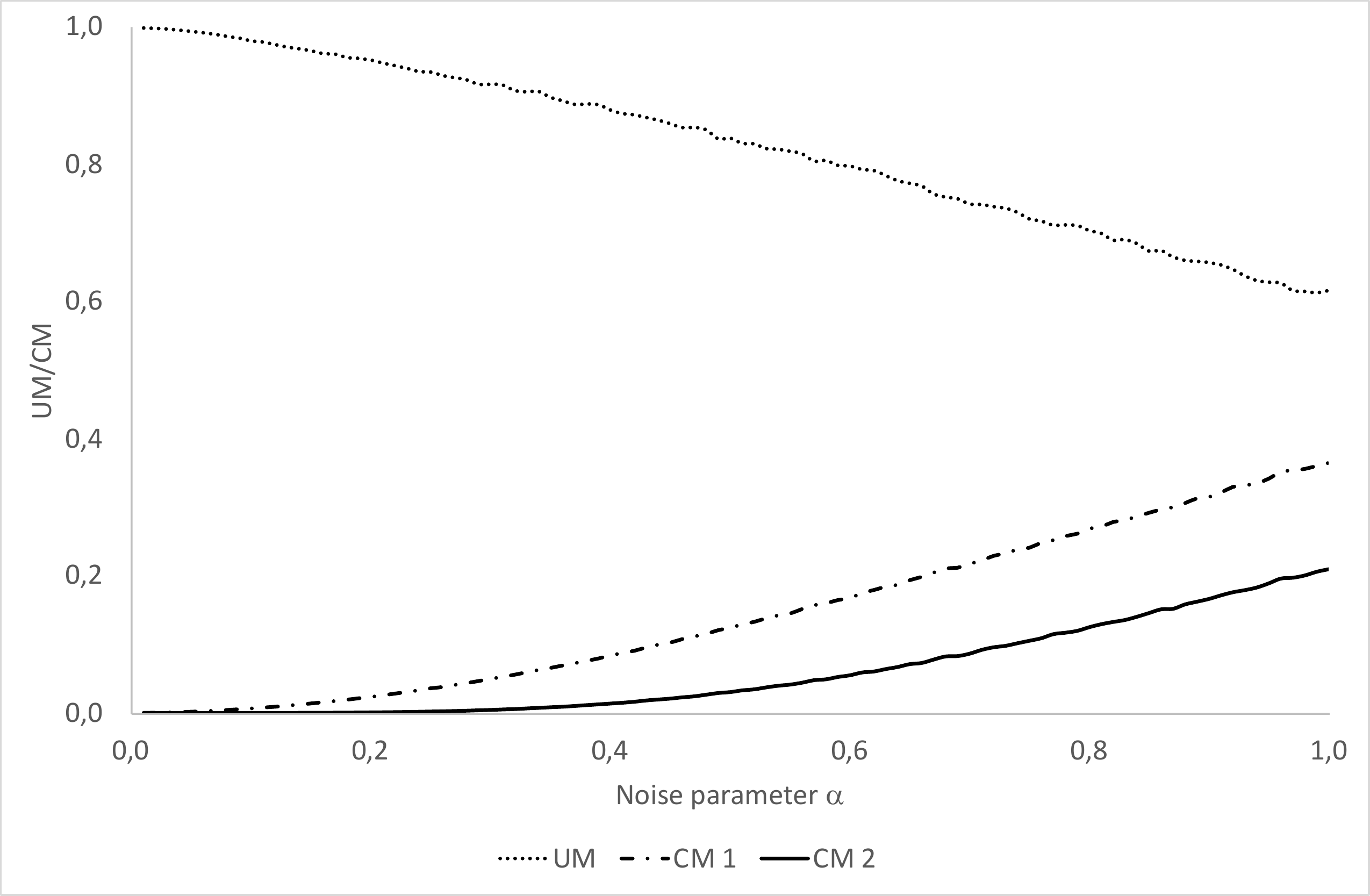}
\end{center}
\caption{Confidentiality and utility metrics for additive noise
as a function of the noise parameter $\alpha$ (the larger $\alpha$,
the larger the noise standard deviation)}
\label{add}
\end{figure*}

\begin{figure*}
\begin{center}
\includegraphics[width=0.7\textwidth]{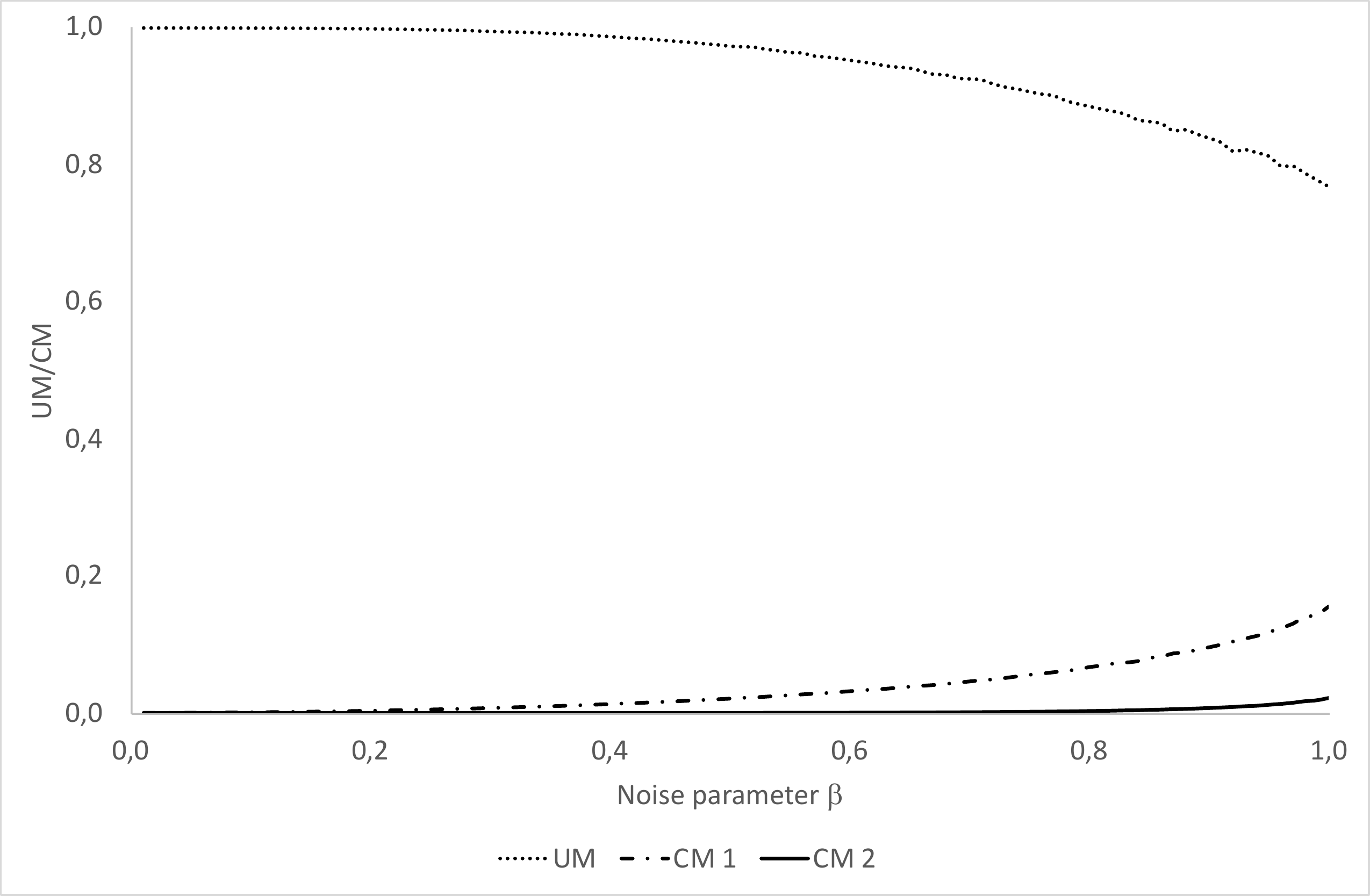}
\end{center}
\caption{Confidentiality and utility metrics for multiplicative
noise as a function of the noise parameter $\beta$ 
(the larger $\beta$, the larger the noise standard deviation)} 
\label{mult}
\end{figure*}

We further tried our metrics on two typical ways of utilizing
noise for statistical disclosure control: noise addition 
%JOSEP. IMPORTANT. Krish to be more specific regarding noise he 
%has used.
and noise multiplication~\cite{Hundepool12}. The 
results are shown in Figures~\ref{add} and~\ref{mult}.
In the former,
an anonymized attribute $Y$ is obtained as 
$Y= X+ E_X$, where $X$ is the corresponding original attribute
and $E_X$ is a noise random variable distributed as $N(0,\alpha \sigma_X)$,
with $0<\alpha \leq 1$ and $\sigma_X$ the standard deviation of $X$.
In multiplicative noise, $Y$ is obtained as 
$Y= X E_X$, where $X$ is the corresponding original attribute
and $E_X$ is a noise random variable generated 
from $Uniform(1-\beta, 1+\beta)$, 
with $0 \leq \beta < 1$.
As it could be expected, it can be seen that, as the noise standard deviation
increases, the utility metric decreases and 
the confidentiality metrics increase. 
For additive noise, these effects are more pronounced: the reason
is that for multiplicative noise the changes in the ranks are smaller
than for additive noise.
Like above, we did not use $CM3$ because in noise addition and multiplication
the controller knows the mapping between original and anonymized records.

\subsection{Synthetic data}
\label{empiricalIPSO}

Finally, we have tried $CM3$ on synthetic 
data generated using the IPSO method~\cite{burridge}. 
IPSO generates a synthetic data set with {\em exactly}
the same means and covariances as the original data set.

The first row of Table~\ref{IPSO} shows the confidentiality metrics
$CM1$, $CM2$ and $CM3$ as well as the utility metric
$UM$ achieved by IPSO when run on ``Census'' as the 
original data set (average of 100 replications). 
The results tell that the synthetic data have high utility,
which was to be expected because IPSO preserves covariances.
Regarding confidentiality, both $CM1$ and $CM2$ give very 
high values, but $CM3$ is substantially lower because
it explores all possible mappings. 
%JOSEP4. Added.
This confirms 
 what we said above: for synthetic data, $CM1$ and $CM2$
should not be used, as they need a mapping and no true
mapping is known.

The second row of Table~\ref{IPSO} shows the metrics 
when IPSO is run on a simulated data set also with 1,080 records
and four attributes, but with very high correlation (0.99) between
the attributes. Like for ``Census'', the results
are the average of 100 replications of IPSO. 
The synthetic data provide very high utility $UM$,
as expected, but the confidentiality according to $CM3$ is very low. 
Note that $CM1$ and $CM2$ are high because they are misled
by the lack of a natural mapping in synthetic data. Thus, 
 $CM3$ is the only confidentiality metric that detects how 
high is the risk of disclosure
in IPSO (or in any other synthetic data generation method)
 in the case of highly correlated original attributes.

\begin{table}
\begin{center}
\caption{Confidentiality and utility metrics 
for synthetic data generated with the IPSO method on
two original data sets: ``Census'' (first row) and a 
simulated data set with highly correlated attributes
(second row)}
\label{IPSO}
\begin{tabular}{|l|c|c|c|c|}\hline
Data set & $UM$ & $CM1$ & $CM2$ & $CM3$ \\ \hline\hline
Census & 0.9638 & 0.9904 & 0.9849 & 0.6673 \\ \hline
Simulated & 1.0000 & 0.9914 & 0.9913 & 0.0277\\\hline
\end{tabular}
\end{center}
\end{table}

\subsection{Summary of experimental results}

%Some of the results of the above experiments  
% might not 
%be generalizable to data sets, SDC methods or parameter
%choices other than those considered: for example, 
%the result that $k$-anonymity yields a slightly better 
%confidentiality-utility trade-off than differential 
%privacy, or the result that, given a confidentiality 
%level, synthetic data yield 
%the highest utility.
%
%However, the above empirical work does illustrate
%two general facts:

The confidentiality and utility measures proposed in this paper are
influenced by the characteristics of the data set. Hence, the results
of the above experiments are not generalizable to data sets,
SDC methods, or parameter choices other than those considered.
However, the above empirical work does illustrate two general facts:

\begin{itemize}
\item The application of the proposed confidentiality and utility 
metrics to substantially different 
privacy models and SDC methods shows that the metrics 
are consistent.
If the confidentiality parameters of the models and/or
methods are set for higher confidentiality (higher $k$ for 
$k$-anonymity, lower $\epsilon$ for differential privacy, higher
noise standard deviation for noise methods), our metrics
detect more confidentiality and less utility. And conversely
if parameters are set for lower confidentiality.
\item Our metrics have been shown helpful to compare 
not only how different parameter values affect the 
confidentiality-utility trade-off of a certain privacy
model or SDC method, but even more interestingly, to 
compare the trade-offs across different privacy models
and/or SDC methods.
\end{itemize}

\section{Related work}
\label{related}

Given that the specific analyses data users will perform on anonymized data
are seldom known by the data protector at the time of anonymization, there 
has been a sustained interest in the literature on generic utility metrics.
On the other hand, there has also been substantial activity to design
confidentiality metrics that could circumvent the costly empirical approach
based on record linkage.
 
In~\cite{Domingo-Ferrer01}, a score was proposed that combines 
utility loss and disclosure risk (confidentiality loss) metrics. 
The approach to disclosure risk 
assessment in that paper relies on record linkage experiments. 
On the other hand, utility loss is measured
by comparing records and some statistics  in the original data set and the 
anonymized data set. Specifically, the mean square error, the mean absolute 
error and the mean variation are used as comparison criteria. The resulting 
utility loss measures are unbounded and thus hard to compare with 
disclosure risk.

In~\cite{Mateo}, a bounded utility loss metric based on probabilities
is presented. 
The metric is the probability that the absolute value of the discrepancy between
a sample statistic $\hat{\Theta}$ and the corresponding population parameter 
$\theta$ is less than
or equal to the discrepancy $|\hat{\theta}-\theta|$ 
measured in the anonymized data set. The intuition is that,
the more different from $\theta$ is the value $\hat{\theta}$ of the sample statistic
in the anonymized data set, the more utility is lost when publishing the 
anonymized data set.
Being bounded, this metric can be readily compared with the risk 
of disclosure, that cannot be above 100\%. 
However, it has the drawbacks of being only applicable to continuous microdata 
and not benefiting from the generality offered by the permutation model.

In~\cite{Oganian}, another generic utility loss metric is proposed that
relies on propensity scores. The original microdata and the anonymized microdata
are merged and a binary attribute $T$ is added that takes value 0 for the original
records and value 1 for the anonymized records. Then $T$ is regressed on the rest
of attributes. Let $\hat{T}$ be the adjusted attribute and 
let the propensity score $\hat{p}_i$ of record $i$ of the merged 
data be the value of $\hat{T}$ for record $i$. Then utility is high if the 
propensity scores of the anonymized and the original records are similar.
This metric is attractive because it focuses on the actual microdata rather
than on preselected statistics. However, it has the drawbacks of being unbounded
and being dependent on the specific regression model chosen. 

In~\cite{NicolasInfSciences}, power means were used
to obtain confidentiality and utility 
metrics based on the permutation model.
The idea is to aggregate the absolute permutation distances $p_1, \ldots, p_n$
resulting from anonymizing the values of an attribute
in the $n$ records of a data set:
\begin{equation}
\label{powermean}
 J((p_1,\ldots, p_n),\alpha) = \left\{ \begin{array}{ll} 
\left(\frac{1}{n}\sum_{i=1}^n p^{\alpha}_i \right)^\frac{1}{\alpha} & 
\mbox{for $\alpha \neq 0$;}\\
\Pi_{i=1}^n p_i^{\frac{1}{n}} & \mbox{for $\alpha=0$,}
\end{array}\right. 
\end{equation}
where $\alpha < 1$ turns the above expression 
into a disclosure risk metric and $\alpha > 1$ into a
utility loss metric.
Indeed, the more $\alpha$ approaches $-\infty$, the greater
is the weight of smaller permutation distances
in Expression (\ref{powermean}); since disclosure occurs
when permutation distances for some values are too small,
we have a disclosure risk metric when $\alpha$ is small.
On the other hand, the more $\alpha$ approaches
$+\infty$, the greater is the weight of larger permutation
distances in Expression (\ref{powermean}); since large
permutation distances are the ones that most deteriorate
utility, we have a utility loss metric when
$\alpha$ is large.
Thus, for $\alpha < 1$, the greater the value 
of $J((p_1,\ldots, p_n),\alpha)$, the more disclosure risk,
whereas, for $\alpha > 1$, the greater the value
of $J((p_1,\ldots, p_n),\alpha)$, the more utility loss.

These power-means metrics may be used
to compare the disclosure protection and the information
loss achieved by two different anonymization methods $M$ and $M'$
(or by the same method $M$ with different parameters $parms$
and $parms'$). However, they have the shortcomings of being intrinsically 
univariate (they operate independently for each attribute) and 
unbounded. In contrast, in this paper, we have proposed
bounded metrics that take all attributes of the data set into account.

\section{Conclusions and future work}
\label{conclusions}

The permutation model is useful to capture the underlying nature of microdata anonymization,
which turns out to be essentially permutation (altering ranks) plus some residual noise 
(to alter values and make them different from the original ones). It seems natural
to leverage this general model to derive general metrics for utility loss
and disclosure risk. This is what we have done in this paper, with the additional 
feature of providing bounded metrics that allow easily evaluating the trade-off
between utility loss and disclosure risk for any anonymization method.

We have presented experimental work that shows that our metrics provide results
that are consistent with the intuition for many anonymization 
approaches in the literature, including privacy models as well
as SDC methods based on noise and synthetic data.
In particular, we have been able to compare the utility-confidentiality
trade-offs achieved by these widely heterogeneous methods, which would
not be possible without the confidentiality and utility methods developed
in this study.
% and we have concluded that
%for the data set considered $k$-anonymity behaves a little better.
%Furthermore, we have evaluated the behavior of the proposed
% metrics when anonymization is 
%based on noise addition and synthetic data generation.
%Comparing these widely divergent approaches would not be possible
%without the utility and confidentiality metrics developed in this study.

Future research lines may include 
comparing the results of our metrics with those obtained with the 
 alternative metrics in the literature mentioned in Section~\ref{related}. 
Also, it may be interesting to compare the confidentiality metrics
with the risk estimated via record linkage and the utility loss
metric with the utility for specific data uses. 
%\item Generalizing the proposed metrics for non-ordinal, that is, nominal 
%categorical 
%attributes. A possibility is to assign to 
%each nominal category a numerical value derived from its semantic
%distance within a certain ontology, as done in~\cite{rufian}. 
%Then an extension of canonical correlations and covariances ought
%to be designed that allows adapting our confidentiality and 
%utility metrics.
%\end{itemize}

\section*{Acknowledgments}

Thanks go to Sergio Mart\'{\i}nez for generating the $k$-anonymous
and the differentially private data sets used in the empirical section.
Partial support for this work has been received 
from the Government of Catalonia 
(ICREA Acad\`emia Award to J. Domingo-Ferrer
and grant 2017 SGR 705), the European Commission (project
 H2020-871042 ``SoBigData++'') 
%JOSEP3. Updated ack.
and the Spanish Government (project RTI2018-095094-B-C21 
``CONSENT''). J. Domingo-Ferrer holds the UNESCO Chair in Data Privacy, but the opinions in
this paper do not commit UNESCO.

\begin{IEEEbiography}[{\includegraphics[width=1in,height=1.25in,clip,keepaspectratio]{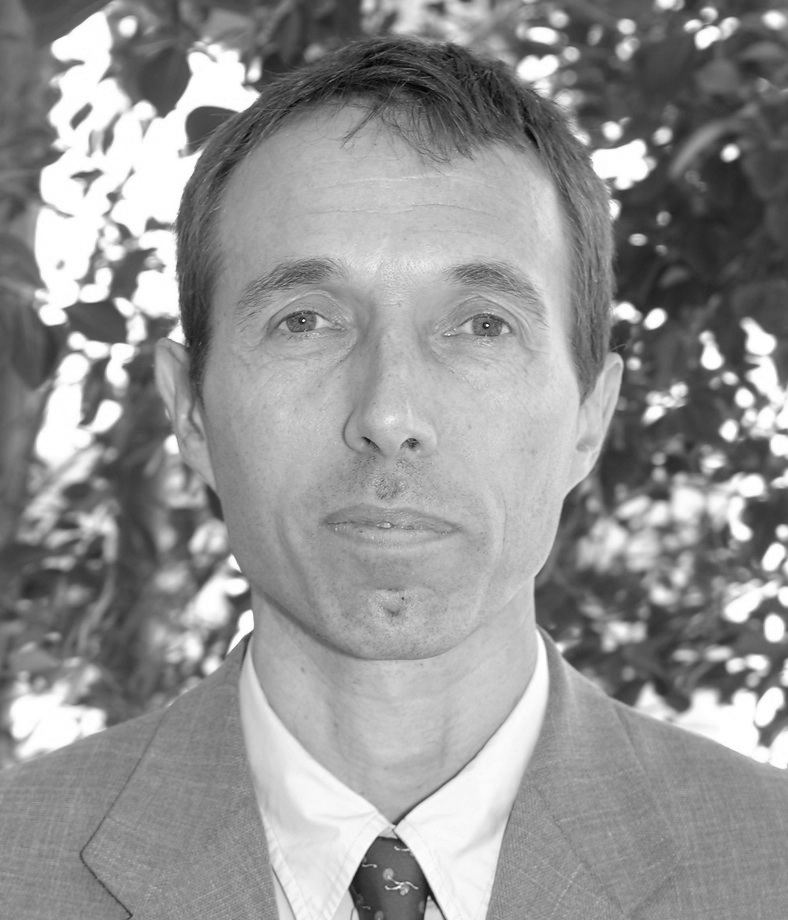}}]{Josep Domingo-Ferrer}
(Fellow, IEEE)
is a Distinguished Professor of Computer Science and
an ICREA-Acad\`emia Researcher at Universitat Rovira i Virgili,
Tarragona, Catalonia, where he holds the UNESCO Chair in Data Privacy
and leads CYBERCAT. He received the MSc and
PhD degrees in Computer Science from
the Autonomous University of Barcelona in 1988 and
1991, respectively. He also holds an MSc degree in
Mathematics.
His research interests are in data privacy, data security and cryptographic
protocols. More information on him can be found
at \url{http://crises-deim.urv.cat/jdomingo}
\end{IEEEbiography}

\begin{IEEEbiography}[{\includegraphics[width=1in,height=1.25in,clip,keepaspectratio]{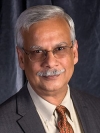}}]{Krishnamurty Muralidhar}
is the Baldwin Chair in Business 
and the Research Director for the Center for Business of Healthcare
at the University of Oklahoma. 
He received a BSc from the University of Madras, India, 
an MBA from Sam Houston State University and 
a PhD from Texas A\&M University.
His main research interests are in data privacy.
\end{IEEEbiography}

\begin{IEEEbiography}[{\includegraphics[width=1in,height=1.25in,clip,keepaspectratio]{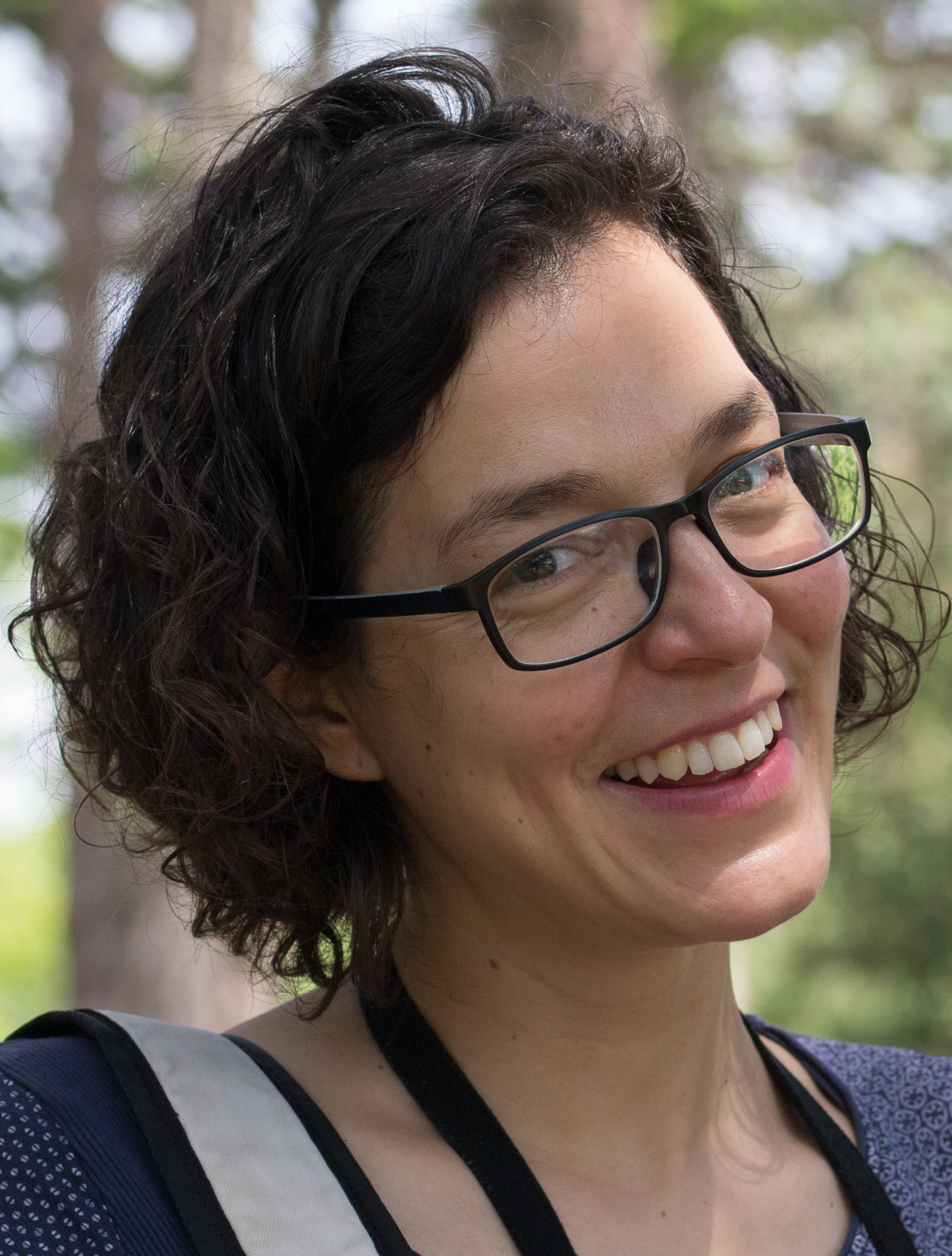}}]{Maria Bras-Amor\'os}
is an Associate Professor of Applied Mathematics 
at Universitat Rovira i Virgili. She received 
her MSc (1998) and PhD (2003) in Mathematics from the Technical 
University of Catalonia (UPC). 
Her main research interests are in data privacy, coding theory
and combinatorics.
\end{IEEEbiography}

\end{document}